\documentclass{aa}
\usepackage{ulem}
\usepackage{graphicx} 
\usepackage{caption}
\usepackage{txfonts}
\usepackage{upgreek}
\usepackage{mathtools}
\usepackage{bm}
\usepackage{subcaption}
\usepackage{comment}
\graphicspath{./images}

\newcommand{\derivp} [2] {\frac {\partial #1 } {\partial #2} }

\newcommand{\algn} [1] {
\begin{align} #1
\end{align}}
\usepackage{xcolor}
\newcommand{\anna}[1]{{#1}}

\title{
Magnetic field morphologies in convective zones influenced by a turbulent surface layer
}
\author{Anna Guseva\inst{1}, Ludovic Petitdemange\inst{1}, Charly Pin\c{c}on\inst{2}}
 \institute{LIRA, Observatoire de Paris, PSL Research University, CNRS, Sorbonne Universit\'e, Paris, France.\\
              \email{anna.guseva@obspm.fr}\\
        \and Université Paris-Saclay, CNRS, Institut d’Astrophysique Spatiale, 91405 Orsay, France\\
        }

\date{\today}

\abstract
{Magnetism of low-mass stars can have a significant impact on their activity, and therefore detection of exoplanets and their properties. Spectropolarimetric observations show that many low-mass stars possess large-scale poloidal magnetic fields with considerable dipole component, which in some cases exhibit temporal dynamics - cycles or reversals. Although it is widely accepted that magnetic fields of low-mass stars are generated by the dynamo process - stretching and twisting of magnetic field lines by helical motions in stellar convective envelopes, numerical dynamo simulations show that it is hard to  reproduce coherent oscillations of large-scale magnetic fields with a dipolar symmetry as observed for the Sun when turbulent and compressible regimes are explored.}
{ Modeling stellar dynamos is a real challenge, as it requires taking into account various interacting physical effects that develop on different time and length scales. Most previous 3D numerical studies partially avoided this problem by considering a numerical domain with low density stratification, which may correspond to neglecting surface effects where density drops considerably.  Our work aims to address this question.
}
{We perform three-dimensional direct numerical simulations of convective dynamos in extreme parameter regimes of both strong turbulence and strong density stratification,  using community-tested numerical software MagIC. The dynamics in such systems, in particular, the dominance of Coriolis effects, depend on the depth.  Our strongly stratified dynamo simulations exhibit rotationally-influenced large-scale convective motions surrounded by a turbulent compressible surface layer.  }
{We find complex time variations of the magnetic field with flow regimes of predominantly dipolar configuration with respect to the few large-scale harmonics as would be captured by spectropolarimetry. In such regimes,  turbulent surface layer induces global magnetic pumping mechanism, transporting magnetic energy into the deep interiors of our dynamo model. Dipole magnetic fields are found in regimes of transition between solar- and anti-solar differential rotation, and  interact dynamically with it.}
{The spatial distribution and temporal behavior of the large-scale fields is consistent with observations of low-mass stars, which suggest magnetic pumping  could promote time-dependent magnetic fields with a dipolar symmetry as observed for the Sun and other solar-like stars. Our results suggest a parameter path in which dynamo models with a complex multiscale dynamics should be explored.}


\titlerunning{Rapid time reversals of dynamo-generated dipolar fields}
\authorrunning{A.Guseva et al.}

\begin{document}

\maketitle
\nolinenumbers
\section{Introduction}

The magnetic activity of the Sun and solar-like stars has captivated astronomers for over a century, beginning with the discovery of sunspot magnetic fields by George Ellery Hale in 1908. These magnetic phenomena, which include the 11-year solar cycle, auroras, and solar flares, arise from the dynamo mechanism within  stellar convection zones \citep{moffatt1978field,dormy2007mathematical,charbonneau2020dynamo,brun2017magnetism}. In such dynamos, the interaction of turbulent convection, differential rotation, and magnetic field generation creates a dynamic and self-sustaining system of magnetic energy with a distribution depending on the astrophysical objects under consideration. Extending our understanding to other stars of similar types reveals both universal principles and fascinating differences in magnetic behaviors \citep{donati2009magnetic,jeffers2023stellar}.

Solar-like dynamos are particularly challenging to model due to the vast range of scales and physical effects involved. From the millisecond timescales of Alfvén waves to the billion-year lifespans of main-sequence stars, a comprehensive model must reconcile processes occurring at vastly different temporal and spatial resolutions. Observational data such as helioseismic profiles provide valuable constraints for the dynamo models but simultaneously reveal the intricacies of phenomena such as differential rotation and meridional circulation that need to be reproduced. Numerical simulations have made significant strides in capturing the broad features of solar-like dynamos, yet they remain limited by computational resources and the inherent complexity of the system \citep{kapyla2023simulations}. In general, dynamo action requires a complex three-dimensional flow structure in order to bypass anti-dynamo theorems \citep{cowling1933magnetic}. Thus, in addition to temporal challenges, modeling stellar magnetism involves bridging length scales ranging from the millimeter-scale eddies of convection to the global-scale flows spanning the entire stellar radius and influenced by rotation and the Coriolis force. The equations governing these magnetohydrodynamic (MHD) systems include nonlinear interactions between these scales that are sensitive to initial and boundary conditions.  Dimensionless parameters like the Rossby number, quantifying rotational influence on convective turbulence, help to categorize different dynamo behaviors with respect to the flow regimes \citep{christensen2006scaling,schrinner2012dipole,gastine2012dipolar,raynaud2015dipolar,menu2020magnetic}. Yet they emphasize the transitions between flow regimes that are still not well understood in the context of stellar physics, such as from solar-like differential rotation with fast equator and slow poles to inverse, anti-solar rotation \citep{gastine2014explaining,karak2015magnetically}. These large-scale zonal flows, generated by convective motions in rotating spherical shells \citep{christensen2001zonal}, are important for the dynamo process as they shear poloidal magnetic field into toroidal field lines (so-called $\Omega$-effect). 

Despite these challenges, advances in observational techniques, such as spectropolarimetric Zeeman-Doppler Imaging (ZDI) \citep{donati2009magnetic}, asteroseismology, and  chromospheric Ca II H and K emission lines  \citep{bohm2007chromospheric}, revealed active and inactive branches of magnetic activity cycles across different types of stars \citep{brandenburg2017evolution}. Observations show a negative correlation between the ratio of magnetic cycle to rotation periods and the Rossby number, at least on the inactive branch, and existence of modulated multi-period cycles, which have been also reported numerically \citep{strugarek2018sensitivity,brun2022powering}.  However,  transitions between these activity branches, as well as the scaling of cycle period with stellar rotation, and the corresponding dynamo mechanisms are still debated~\citep{olspert2018estimating,saikia2018chromospheric}. Further discrepancies remain,  such as the inability of simulations to fully replicate the Sun’s convective amplitudes or equatorward propagation of dynamo waves with a preferred dipolar symmetry with a complex temporal behavior on long timescale \citep{raynaud2016convective}. The magnetic fields resulting from Boussinesq \citep{busse2006parameter,goudard2008relations,schrinner2011oscillatory,schrinner2012dipole} or anelastic models \citep{gastine2012dipolar,schrinner2014topology,strugarek2017reconciling,strugarek2018sensitivity,pinccon2024coriolis} exhibit cyclic magnetic behaviors with low magnetic Reynolds numbers and a parity often quadrupolar whereas, on the other hand, Sun's magnetic field is dominated by its dipolar components that have periodic oscillations with very high magnetic Reynolds numbers.

Surface stellar magnetic fields, reconstructed from inversion of multiple polarized spectral lines over several stellar rotation periods with ZDI, are of particular interest, as they simultaneously infer activity and magnetic topology of stellar dynamos. Examples of stars exhibiting solar-like magnetic cycles with predominantly dipolar fields during low levels of activity include 61 Cyg, $\epsilon$ Eri, $\kappa$ Cet, with activity periods of 7 to 13 years \citep{jeffers2023stellar},  and F-type stars $\tau$ Boo and HD 75332, with shorter cycles of 1-3 years \citep{jeffers2018relation,brown2021magnetic}. On the other hand, younger K stars like EK Dra and HN Peg have rapidly evolving magnetic fields without correlation with S-index. Unexpected, potentially acyclic reversals of magnetic topology or minima in activity were also observed, e.g. in LQ Hya, HD 16620 \citep{willamo2022zeeman,lehtinen2022topological}; during magnetic reversals, the dominant magnetic topology switches to multi-polar or quadrupolar~\citep{saikia2018direct}. These observations suggest that stellar cycles can become intermittent and aperiodic at certain stages of stellar evolution \citep{jeffers2023stellar,van2016weakened,metcalfe2022origin}.
Although surface magnetic fields are influenced by small-scale turbulence, their large-scale components must evolve with the global nature of dynamo action and this coupling represents an important motivation for the actual study. 

In this work, we focus on the dynamical mechanisms of large-scale magnetic behaviour in strongly stratified and turbulent models of low-mass stars, and show that density stratification can promote dipolar, albeit aperiodic dynamo states in such systems. Such stratification separates the flow into inner layer of rotationally constrained convective columns and outer layer of anisotropic turbulent small-scale convection, promoting magnetic pumping of small-scale magnetic fluctuations inside the tangent cylinder (imaginary cylindrical surface around the inner boundary of our computational domain). This process enhances the dipolar magnetic mode, which competes with a more complex multipolar magnetic state and dynamically interacts with differential rotation, preventing transition to anti-solar rotation. This paper is structured as follows. In section~\ref{sec:methods} we briefly discuss the anelastic model and our choice of parameter regime.  In section~\ref{sec:mag_solutions} we describe dynamo states observed in our simulations. Section~\ref{sec:cycle} focuses of the dipole solutions that manifest themselves in strongly stratified convection, and section~\ref{sec:mechanisms} presents the argument for magnetic pumping supporting dipolar solutions. In section~\ref{sec:dics} we discuss of our results in the context of previous work and stellar observations, and conclude our findings in section~\ref{sec:conc}.

\section{Anelastic simulations}\label{sec:methods}
\subsection{Setup}
We perform direct numerical simulations of the dynamo process in anelastic approximation using MAGIC code~\citep{gastine2014explaining}, solving equations for velocity, magnetic fields, and entropy in spherical coordinates $(r, \theta, \phi)$, as described in Appendix~\ref{sec:app_eqn}. Cylindrical coordinates $(r_c, \phi, z_c)$ are also used where necessary. Conducting fluid is contained in a shell between two spheres with radius ratio $\chi = r_i/r_o = 0.35$, co-rotating with angular velocity $\Omega$ and representing a thick convective envelope of a solar-like star. Convection is triggered by the entropy contrast $\Delta S$ fixed at the boundaries of the spherical shell, under assumption of adiabatic background state. Non-uniform density between the two spheres, modeling density decrease with radius in stellar envelopes, is set by imposing background density profile with polytropic index $n=2$.  The key dimensionless parameters explored in our work are Rayleigh number $Ra$ and density contrast $N_\rho$, defined as
\begin{equation}\label{eq:params}
    Ra = \frac{GM d\Delta S}{\nu \kappa c_p}, \quad N_\rho = \ln{\frac{\rho_i}{\rho_o},}
\end{equation}
where $\nu$ is hydrodynamic viscosity, $\kappa$ thermal diffusivity, $c_p$ heat capacity, $d=r_o-r_i$ the gap between  the two spheres,  $G$ gravitational constant, and $M$ the central mass. Rayleigh number determines the strength of convection relative to diffusive transport; convection is driven at a certain critical value  $Ra_{cr}$. This value increases with density contrast \citep{JonesKM09,raynaud2015dipolar}, and so simulations at high $N_\rho$ become challenging. To approximate stellar conditions, it is necessary to set 
\begin{equation}
    \widetilde{Ra} = Ra/Ra_{cr} \gg 1, \quad \rho_i/\rho_o \gg 1.
\end{equation}

The rest of dimensionless parameters governing the system are Ekman number $E=\nu/\Omega d^2$, Prandtl number $Pr =\nu/\kappa$, and magnetic Prandtl number $Pm = \nu/\eta$, where $\eta$ is magnetic diffusivity.  We fixed $E=3 \times 10^{-4}$, corresponding to moderate rotation as compared to viscosity, $Pr=1$ and $Pm=2$. When compared to estimations of these parameters in the stars (e.g. $Ra \sim 10^{20}$, $E \sim 10^{-15}$, $Pr\sim 10^{-7}$, $Pm \sim 10^{-5}$ for the Sun), these values in simulations can be interpreted as if additional turbulent dissipation was present in the simulations.  Note that $\widetilde{Ra}$ is calculated based on the entropy difference across the convective zone, which is hard to measure in stellar or planetary context, and includes dissipative effects in its definition, which are considered weak in this context. A modified flux-based Rayleigh number $Ra^*_F = Ra Nu Ek^3/Pr^2$, eliminating dissipative effects and based on the heat flux that is easier to estimate from observations, has been proposed by \cite{christensen2002zonal} and \cite{aurnou2020connections} as a scaling of convection strength. The Nusselt number $Nu$ is defined as the output luminosity to the basic state luminosity (see appendix~\ref{sec:app_sim_param}). Our data show that flux-based Rayleigh number scales linearly with $\widetilde{Ra}$ (table~\ref{tab:sim_param}); for simplicity, we keep $\widetilde{Ra}$ as a parameter in this work. 

\begin{figure*}[h!]

      \centering        
 \begin{subfigure}[b]{0.32\textwidth}
         \centering
        \caption{}
         \includegraphics[width=\textwidth]{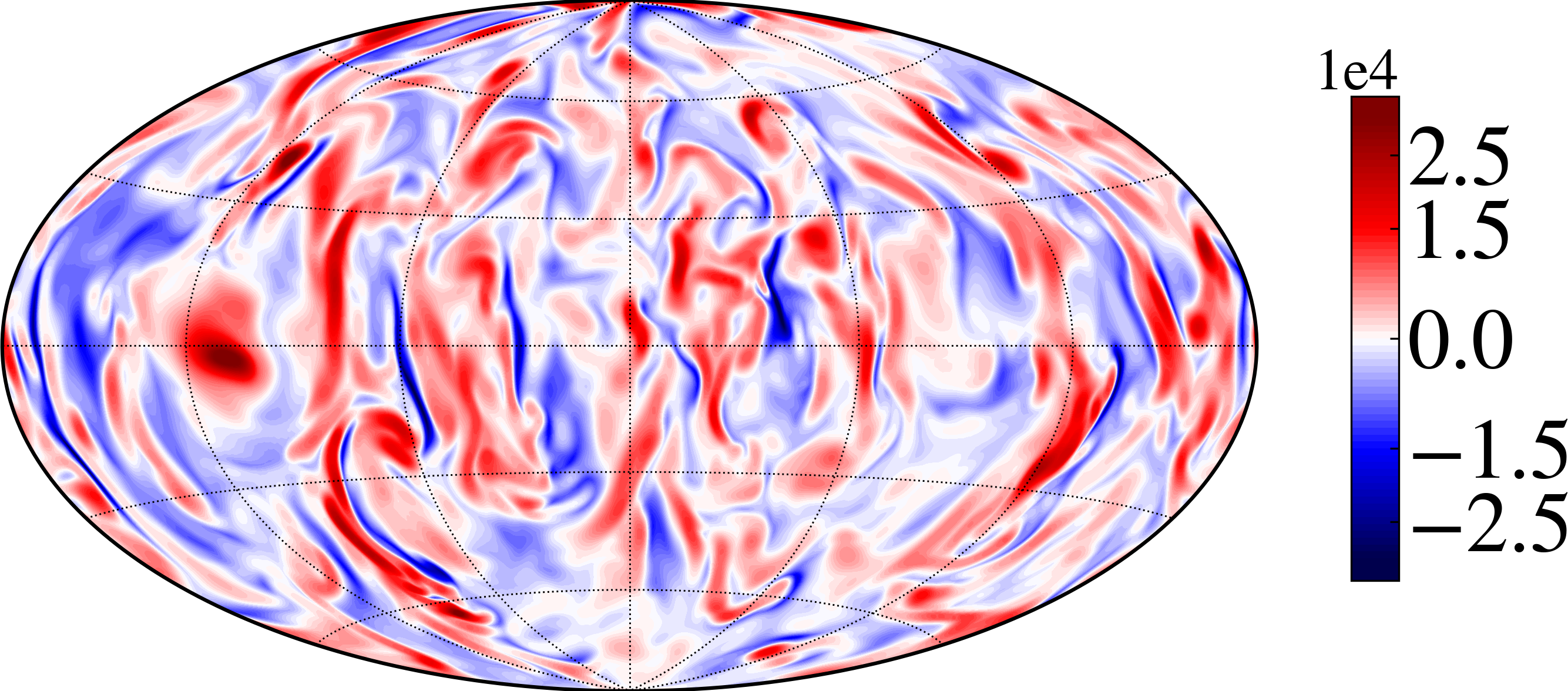}
     \end{subfigure}
 \hfill
     \begin{subfigure}[b]{0.32\textwidth}
         \centering
        \caption{}
         \includegraphics[width=\textwidth]{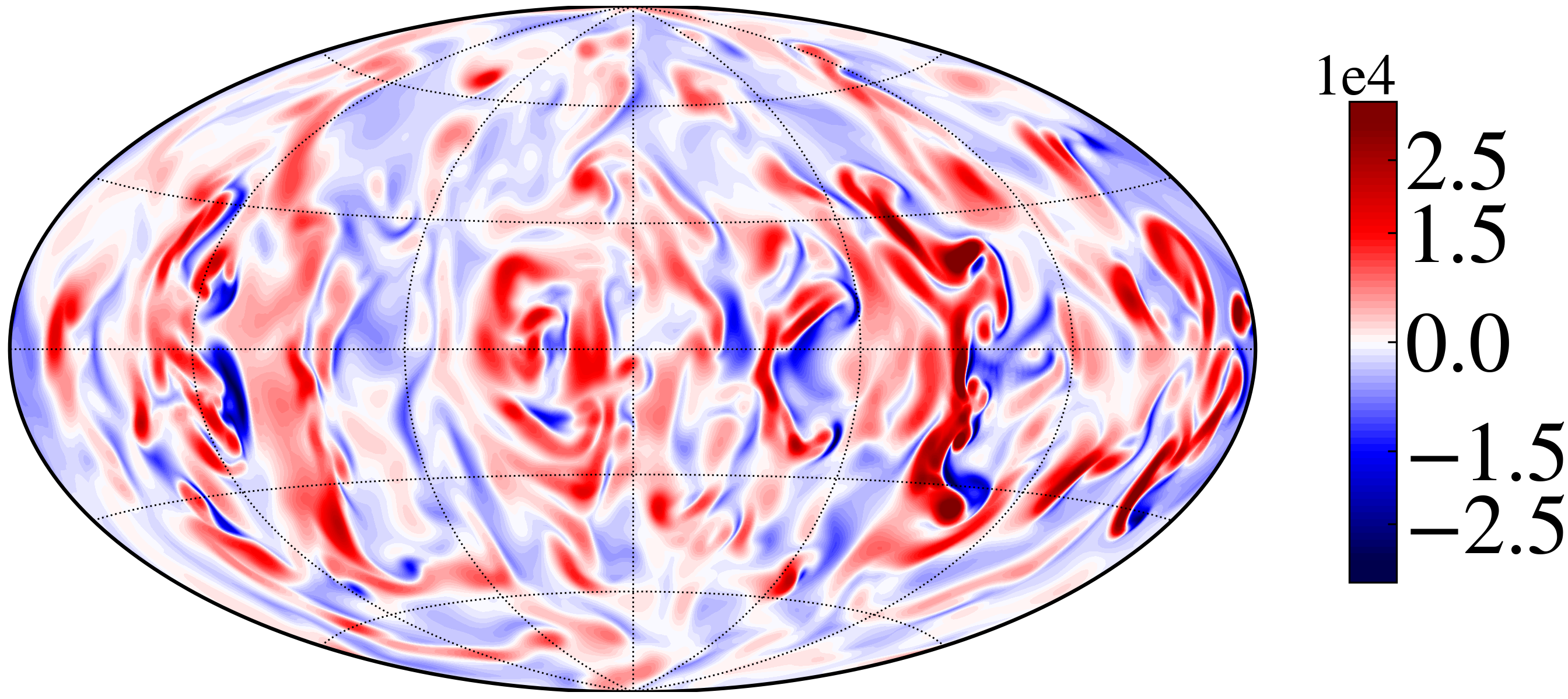}
     \end{subfigure}
      \hfill
     \begin{subfigure}[b]{0.32\textwidth}
         \centering
        \caption{}
         \includegraphics[width=\textwidth]{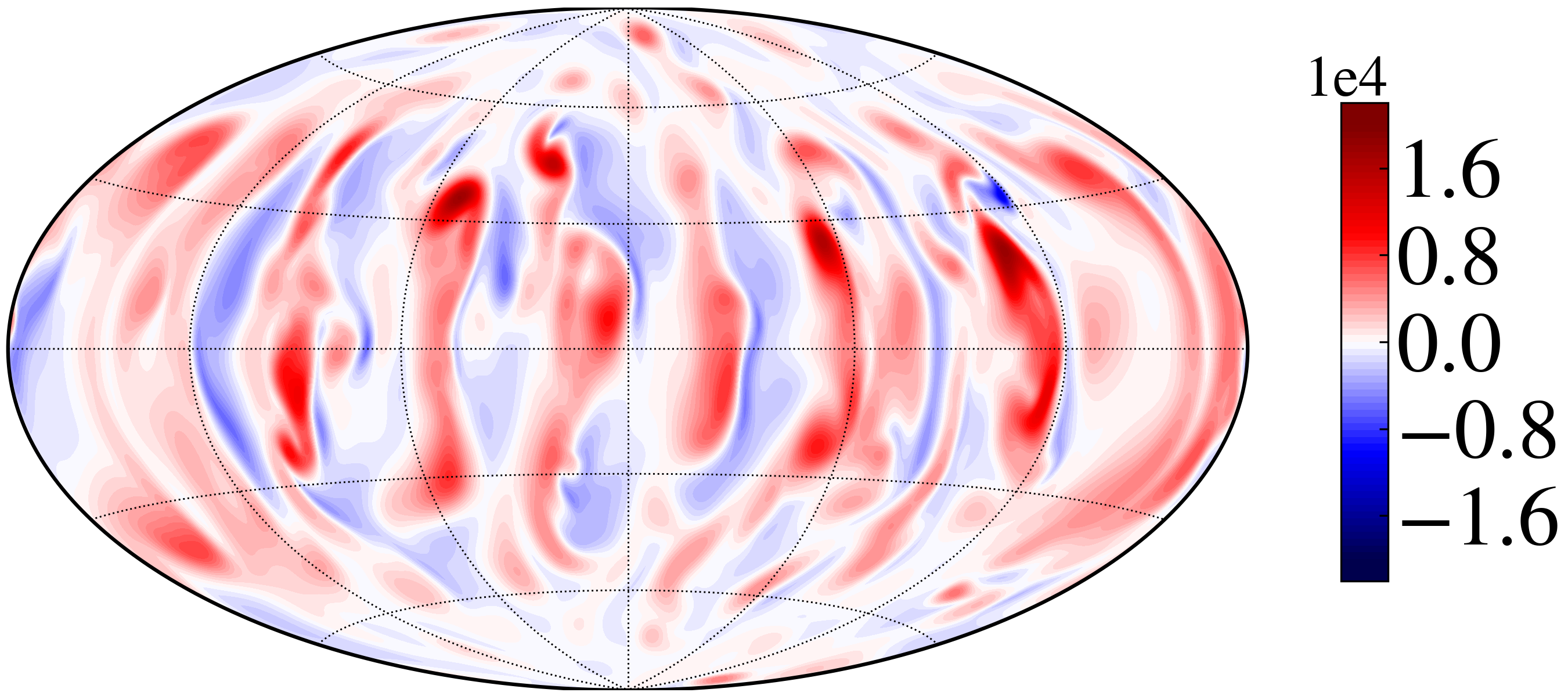}
     \end{subfigure}
     \vfill 
\begin{subfigure}[b]{0.32\textwidth}
         \centering
        \caption{}
         \includegraphics[width=\textwidth]{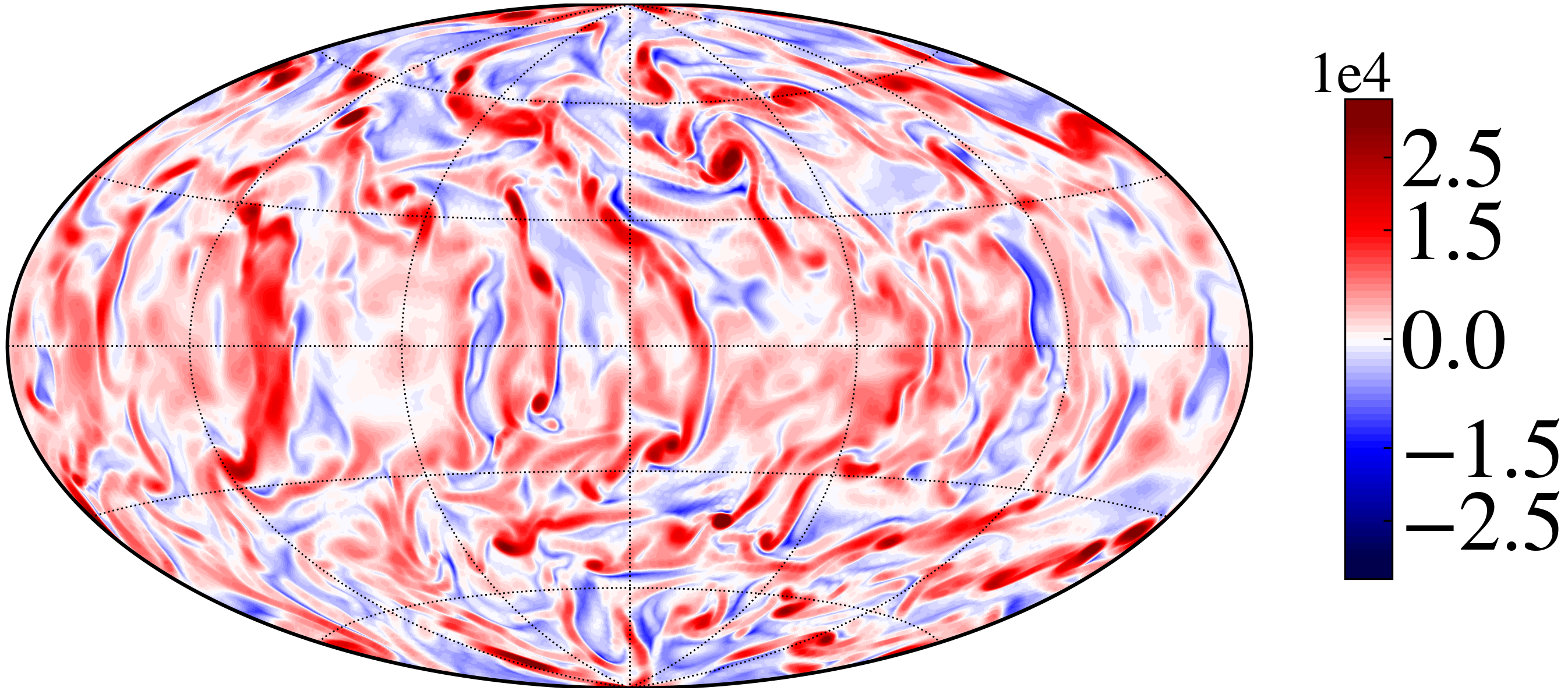}
     \end{subfigure}
 \hfill
 \begin{subfigure}[b]{0.32\textwidth}
         \centering
        \caption{}
         \includegraphics[width=\textwidth]{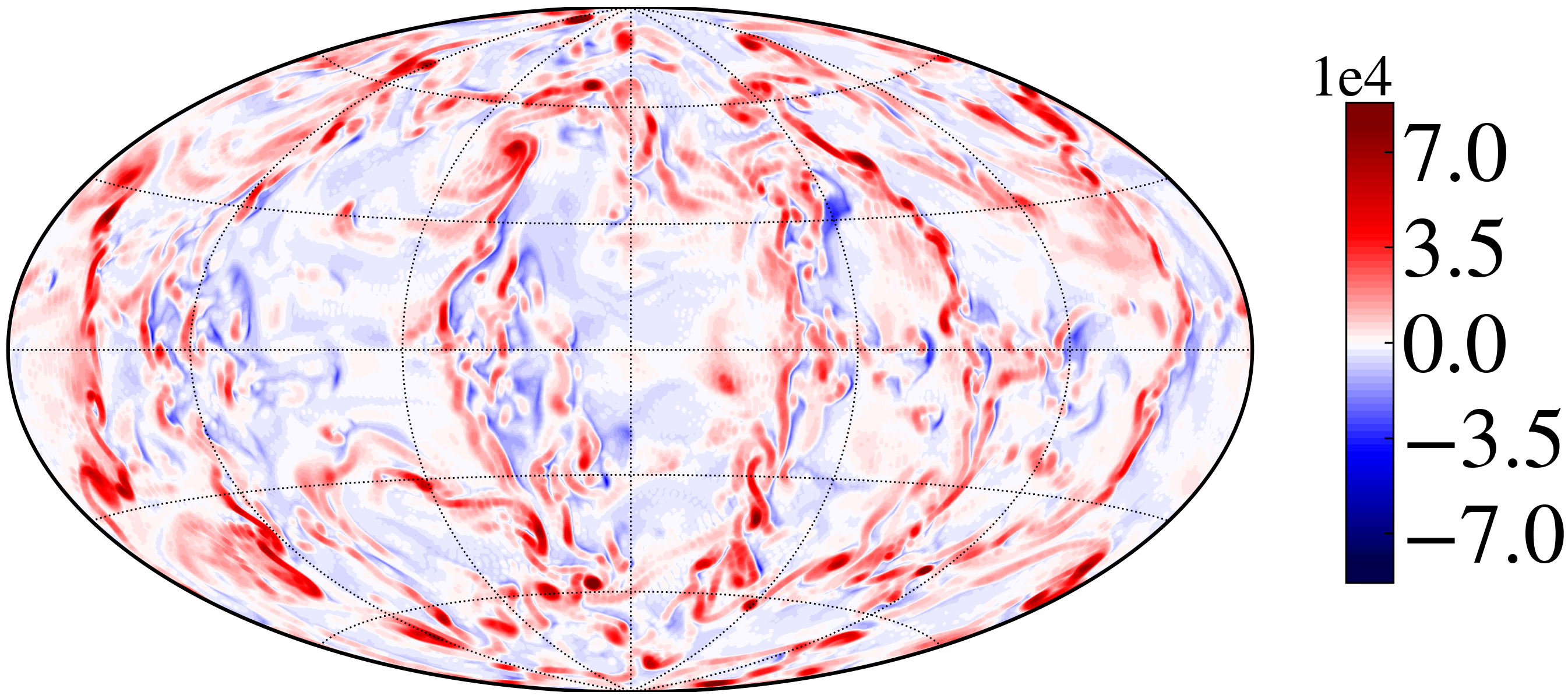}
     \end{subfigure}
     \hfill
   \begin{subfigure}[b]{0.32\textwidth}
         \centering
        \caption{}
         \includegraphics[width=\textwidth]{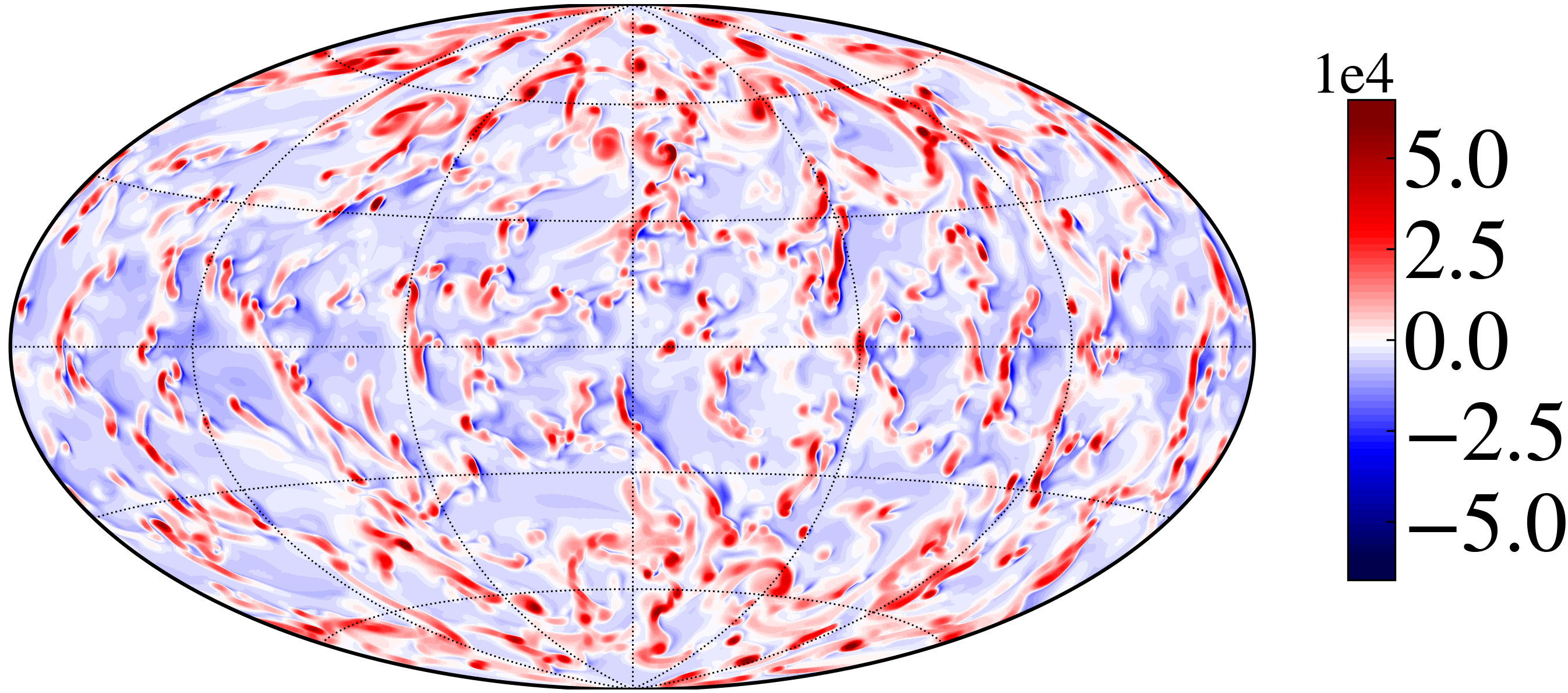}
     \end{subfigure}   
        \caption{ Instantaneous surfaces of vertical vorticity component of the flow at $r/r_0=0.5$ (top) and $r/r_0=0.9$ (bottom), illustrating scale separation developing with $N_\rho$ in strongly turbulent convection, $\widetilde{Ra} =32$. (a,d) $N_\rho=2$; (b,e) $N_\rho=4$; (c,f) $N_\rho=6$. }
        \label{fig:vr_slices}
\end{figure*}

A typical simulation in our work was set up as follows. As an initial condition for magnetic, velocity and temperature fields, we used a steady-state flow snapshots from the convective dynamo database of \cite{pinccon2024coriolis}. Furthermore, we employ stress-free boundary conditions for velocity field and insulating boundary condition for magnetic field. Typical resolution in our simulations span from $N_r \times N_\phi = [65, 576]$ to $[241, 1280]$ where $N_r$ denotes the number of Chebyshev polynomials in radial direction and $N_\phi$ the total number of spherical harmonics. The resolution was set to have  at least three order of magnitude difference between the energy in the largest and the smallest scales in the energy spectra for each combination of $N_\rho$ and $\widetilde{Ra}$, and the simulations were run at high resolution until a steady state was obtained again. After that, the number of spherical harmonics was reduced to speed up the calculation and gather long-term flow statistics and snapshots. For more details on the parameters, flow equations and numerical methods, see appendices~\ref{sec:app_eqn},~\ref{sec:app_sim_param}, and the website of MagIC code\footnote{\url{https://magic-sph.github.io/}}.

\subsection{Choice of parameter regime}\label{sec:conv}

In this section, we explore $N_\rho =(2,4,6)$ corresponding to density ratios of $\rho_i/\rho_o = (7.4,54.6,403.4)$ between the inner and the outer spheres, and provide a motivation for the choice of parameter regimes for our dynamo study. 

Firstly, we compare the flow structures for these three values of $N_\rho$ at the same level of turbulence,  $\widetilde{Ra} =32$. In rotating convection, the Coriolis force tends to remove vertical gradients and align flow structures along the rotation axis. When convective turbulence sets in, inertia forces due to nonlinear interactions of flow fluctuations are in competition with Coriolis force and tend to break this alignment. The result is a rather three-dimensional structure of the flow, as is visible from the surface distribution of vertical component of flow vorticity in the bulk of the domain, $r/r_o=0.5$ for $N_\rho=2$ (figure~\ref{fig:vr_slices}a). Since the density contrast is low in this case, the structure of convection do not vary considerably along the radius, and so distribution of vorticity across different length scales is similar both in the bulk of the domain, $r/r_o=0.5$, and near the surface, $r/r_o = 0.9$ (figure~\ref{fig:vr_slices}a,d). However,  simultaneously increasing convection and stratification has a profound effect on turbulent flow properties: for the same level of convection, its length scale decreases toward the surface for $N_\rho=4$ (figure~\ref{fig:vr_slices}b,e). Further increase of stratification to $N_\rho=6$ results in extremely strong dynamical contrast between the surface and the bulk, where large-scale, rotationally constrained convective columns develop at $r/r_o=0.5$ (figure~\ref{fig:vr_slices}c), and small-scale, three-dimensional vortical structures at the surface (figure~\ref{fig:vr_slices}f). The flow in the domain is essentially separated in a Coriolis-affected interior and a strongly turbulent surface layer. The influence of surface layers on dynamics was often overlooked in the previous works, although such layers possess strong kinetic helicity and are thus natural candidates to generate magnetic fields.

\begin{figure*}
           \centering 

     \begin{subfigure}[b]{0.475\textwidth}
         \centering
        \caption{}
         \includegraphics[width=\textwidth]{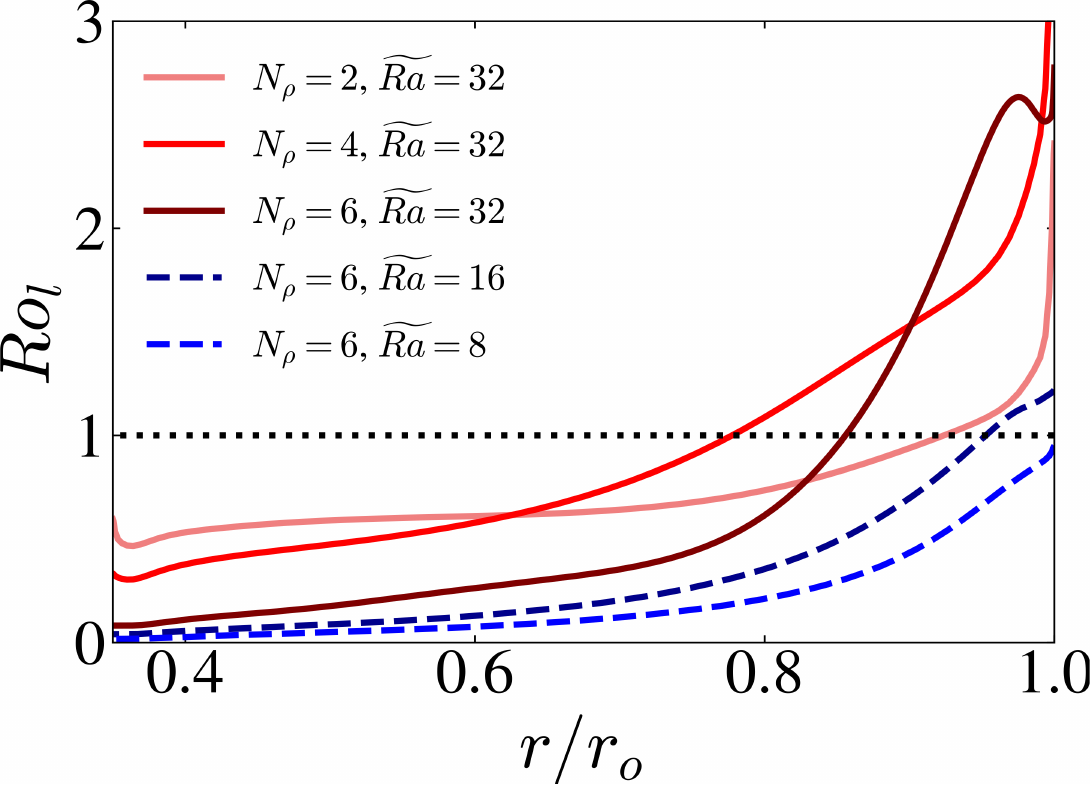}
     \end{subfigure}  
      \hfill
 \begin{subfigure}[b]{0.475\textwidth}
         \centering
        \caption{}
         \includegraphics[width=\textwidth]{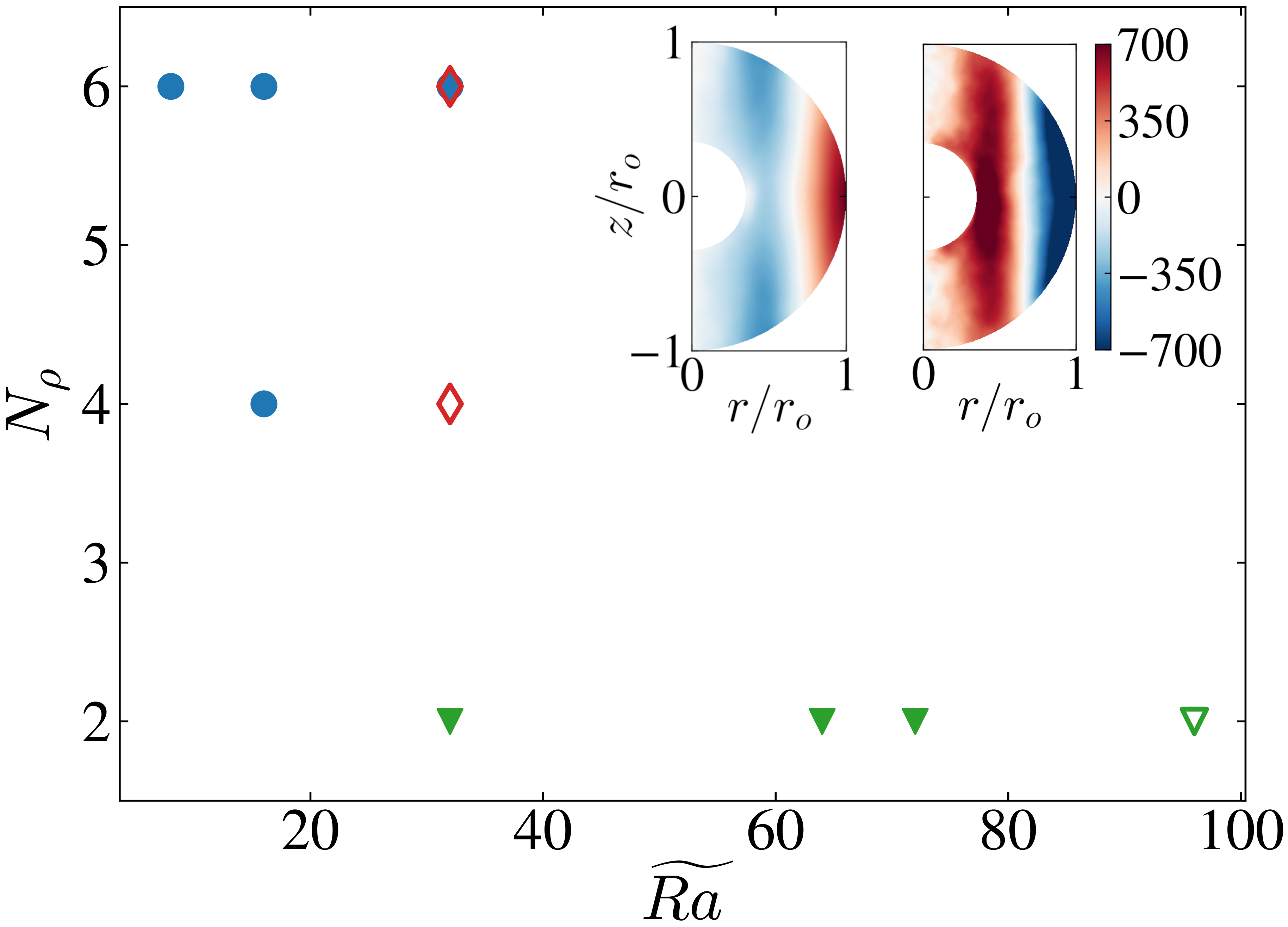}
     \end{subfigure}

        \caption{ (a) Effect of varying $N_\rho$ (in red, $Ra/Ra_{cr} =32$) and $Ra$ (in blue, $N_\rho =6$) on local Rossby number calculated based on convective length scale $l_{conv}$ \eqref{eq:rossby}. Dotted line denotes $Ro_l=1$, e.g. a balance between inertia and Coriolis force. (b)  Density contrasts $N_\rho$ and Rayleigh numbers $\widetilde{Ra}$ spanned by simulations in this work. Different symbols denote three dynamical regimes of the flow, discussed in more details in section~\ref{sec:mag_solutions}. Filled (empty) triangles, predominantly multipolar fields with solar (anti-solar) differential rotation (section~\ref{sec:multi_sol}). Circles, quadrupolar oscillatory dynamos with solar-like differential rotation (section~\ref{sec:waves_sol}). Diamonds, reversing dipole-multipole solutions with transitions between three-layered and anti-solar rotation (section~\ref{sec:dip_sol}). Solar- and anti-solar differential rotation  is shown in the inset plots depicting zonal flows $\langle u_\phi\rangle_\phi$ for $N_\rho=2$, $\widetilde{Ra}=32$ (left) and $\widetilde{Ra}=96$ (right).}
        \label{fig:sim_param}
\end{figure*}

Quantitatively, the separation of scales in stratified convection can be described by the local convective Rossby number, representing the balance between inertia and Coriolis forces at each radii,
\begin{equation}\label{eq:rossby}
    Ro_l (r) = \frac{u_{rms} (r)}{l_{conv} (r) \Omega }, \quad l_{conv} = \frac{ \pi E^{kin} (r)}{\sum_l l E^{kin}_{l} (r)}, \quad m \neq 0, 
\end{equation}
where $u_{rms}$ is the root-mean-square convective velocity and $l_{conv}$ is the typical length scale of convection calculated from the distribution of kinetic energy $E^{kin}$ with spherical harmonic degree $l$ \citep{christensen2006scaling}, \anna{given in the units of $d$}. The energy contained in the axisymmetric, azimuthal zonal flow $\langle u_\phi \rangle_\phi$, averaged over longitude $\phi$, is subtracted to separate proper convective motions from differential rotation driven by them. For stars, it is not straightforward to estimate this local parameter, and so global Rossby number, the ratio of stellar rotation period to empirically inferred convective turnover time, is  used instead \citep{noyes1984relation}.

Figure~\ref{fig:sim_param}a presents the dependence of the local $Ro$ number with radius for varying $N_\rho$ ($\widetilde{Ra}=32$, solid). For $N_\rho=2$, it is relatively large,  $Ro_l\sim 0.5$,  and is nearly constant across the spherical shell. It  increases proportionally in the entire domain with increasing $\widetilde{Ra}$ and enhancement of convection. Increasing density stratification consistently decreases $Ro_l$ in the deep interiors, developing there a rotationally constrained zone $Ro_l \ll 1$; for $N_\rho=6$, it occupies about a half of the entire simulation domain. There, the dominance of the Coriolis force enforce Taylor-Proudman constraint to produce elongated convective columns, as seen in figure~\ref{fig:vr_slices}c; effectively, the flow in the bulk becomes less dependent of the vertical coordinate $z$. On the other hand,  the radial profile of $Ro_l$ becomes much steeper throughout the domain if compared to $N_\rho=2$. The area with $Ro_l>1$, developing for $N_\rho=4$ and $6$, corresponds to the strongly turbulent surface layer with inertia-dominated small-scale vortices (figure~\ref{fig:vr_slices}f). The increase in $Ro_l$ at the surface with $Ra$ (figure~\ref{fig:sim_param}a) is accompanied by a decrease of the length scale of convection  (not shown here), meaning that local $l_{conv}(r)$ at the surface becomes about 5 \anna{times} smaller than in the bulk. This essentially creates a scale separation between convection in the bulk and the surface layer, which can be observed in figure~\ref{fig:vr_slices}(c,f), with large scales much more energetic in the bulk, and small scales at the surface. To develop solutions with such pronounced scale separation,  high values $\widetilde{Ra}$ are needed. Figure~\ref{fig:sim_param}a shows that $Ro_l$ gradually increases with increase of $\widetilde{Ra}$ at the outer shell (dashed lines, fixed $N_\rho=6$), developing a surface area where $Ro>1$.  Thus, it is the simultaneous increase of density stratification and enhancement of convection that  results in Coriolis force dominating in the bulk, and turbulent inertia dominating the outer layer.

Previous studies of non-stratified convection showed that convective turbulence with strong inertia and $Ro_l>1$ develops multi-polar, disorganized dynamo solutions \citep{christensen2006scaling}. Similar behavior was found for weakly stratified anelastic convection~\cite{gastine2012dipolar,raynaud2015dipolar}, who found that the stability range  of dipolar dynamos shrinks in $\widetilde{Ra}$ with increase of $N_\rho$. In this work, we are interested in the joint effect of stratification and turbulence, and scale separation, on the dynamo properties. We thus explore parameter space around solutions with $N_\rho=4,6$ and $\widetilde{Ra}=32$, more representative of stellar flows. The entire set of our simulations, all in fully turbulent convective regimes, is schematically presented in figure~\ref{fig:sim_param}b. Due to numerical constraints, we probe a wider range of turbulent states for $N_\rho=2$, with $\widetilde{Ra} \in [32,96]$,  than for $N_\rho = 6$ ($\widetilde{Ra} \in [8,32]$).

An important feature of this parameter space is transition from solar-like differential rotation, with faster equator and slower poles to anti-solar differential rotation with slower equator and faster poles (see inset plots in figure~\ref{fig:sim_param}b). For $N_\rho=2$, this transition takes place between $\widetilde{Ra}=72$ and $\widetilde{Ra}=80$; our most stratified flows with $\widetilde{Ra}$ appear to be near the threshold of such transition. This  affects the properties of dynamo and convection; for example, the structure of convection is modified by differential rotation for $N_\rho=6$ and $\widetilde{Ra}$, which creates a local increase in convective length scale  near $r_o$ and the corresponding local deficiency in $Ro_l$ (figure~\ref{fig:sim_param}a), comparable to the thickness of weak retrograde zonal jet at the surface (see section~\ref{sec:shear2flux} and figure~\ref{fig:flux_streamlines}c). Note that all simulations discussed below are magnetohydrodynamic dynamos; in the absence of feedback of magnetic stresses on the flow, transition to anti-solar rotation takes place at lower $\widetilde{Ra}$.

\section{Types of dynamo states in simulations}\label{sec:mag_solutions}
In this section, we examine the three different types of dynamo solutions observed in the flow for the set of simulations in figure~\ref{fig:sim_param}b. While all of our simulations are fully turbulent dynamos with energy spread across a range of scales, their dynamics is different if the large scales are considered. To illustrate these differences, we use the following characteristics: butterfly diagrams of the radial component of magnetic field $B_r$, filtered snapshots of the large-scale magnetic field, and zonal flows. In the following, we will discuss in details these diagnostic quantities and resulting dynamo states.
\begin{figure*}[h!]
\centering
    \begin{subfigure}[b]{\textwidth}
         \centering
        \caption{}
         \includegraphics[width=0.8\textwidth]{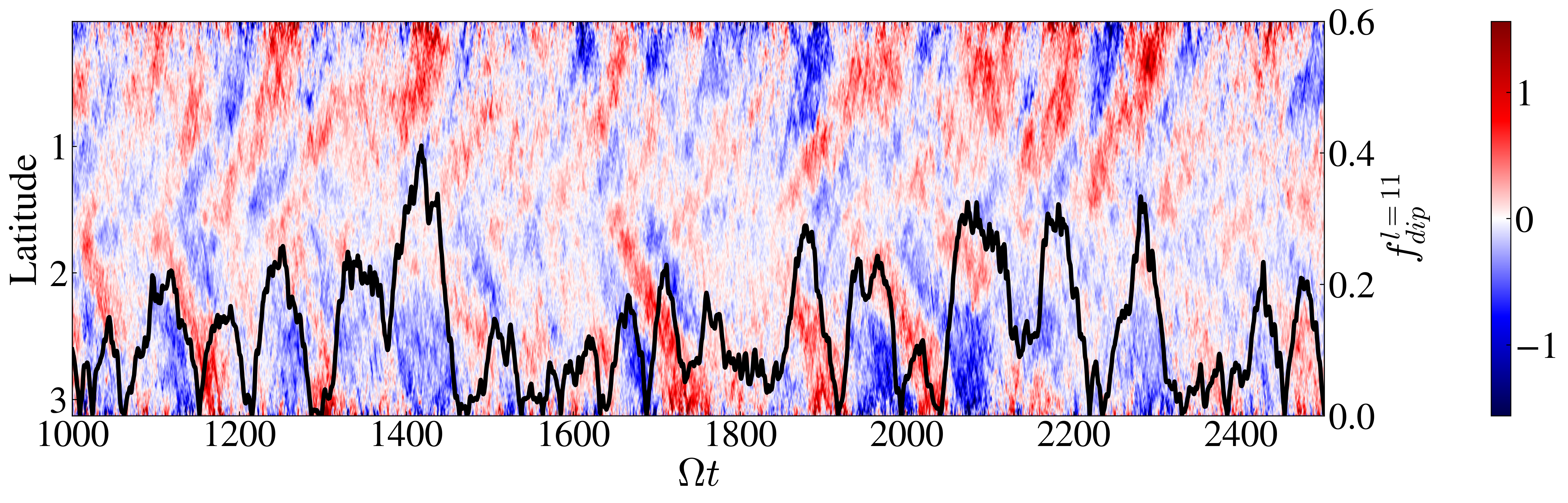}
     \end{subfigure}
     \hfill 
         \begin{subfigure}[b]{0.8\textwidth}
         \centering
        \caption{}
         \includegraphics[width=\textwidth]{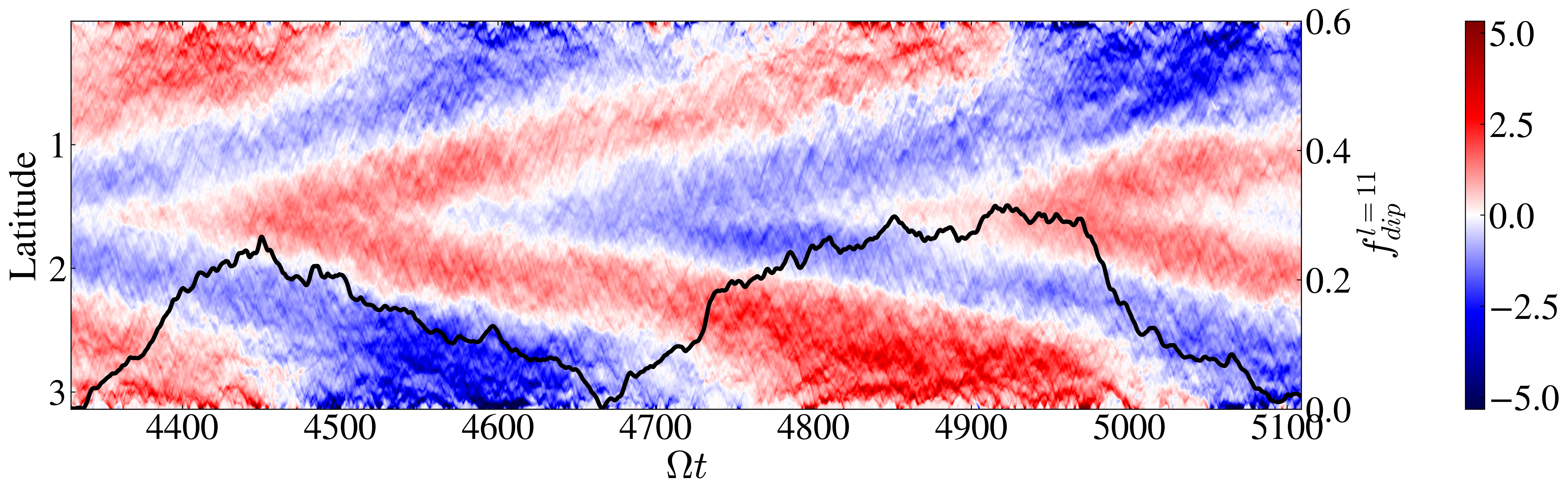}
     \end{subfigure}
     \vfill
         \begin{subfigure}[b]{0.8\textwidth}
         \centering
        \caption{}
         \includegraphics[width=\textwidth]{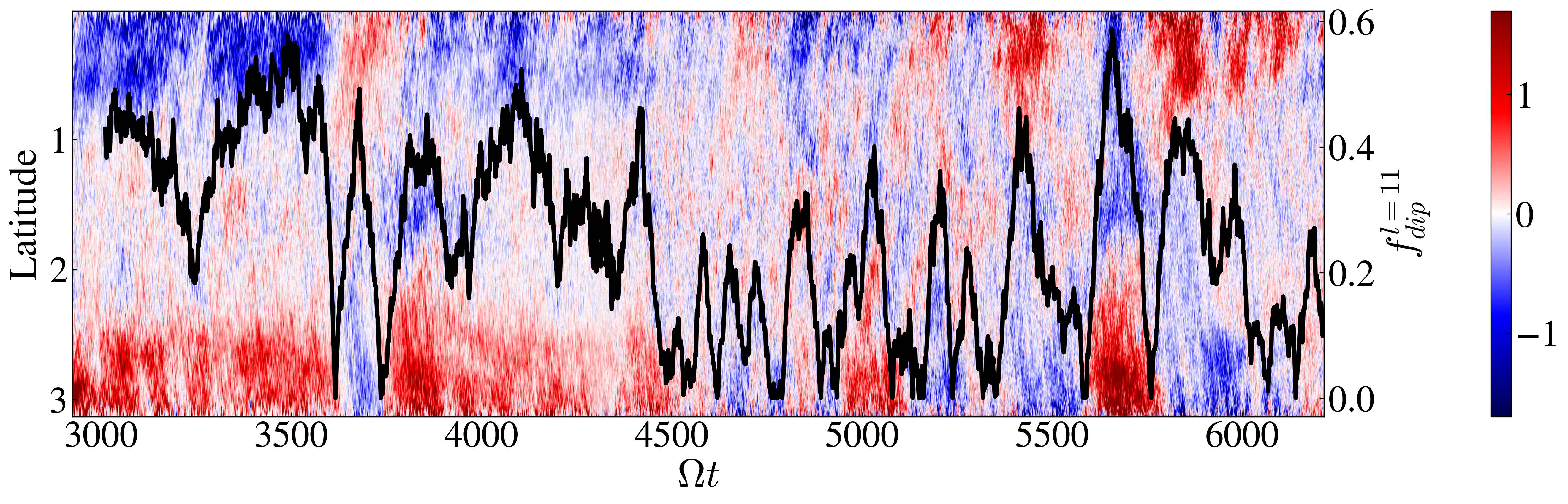}
     \end{subfigure}
             \caption{Butterfly diagrams of axisymmetric, azimuthally-averaged magnetic component $\langle B_r \rangle_\phi$ at the outer boundary. (a)  $N_\rho =2$, $Ra/Ra_{cr}=32$; (b) $N_\rho = 6$, $Ra/Ra_{cr}=16$; (c) $N_\rho =6$, $Ra/Ra_{cr}=32$. To highlight large-scale components, the colormaps were saturated with cut-offs of $0.2$, $0.25$ and $0.1$, respectively; the magnitude of magnetic fluctuations is larger by a corresponding factor. Black curves correspond to the surface dipolarity with cut-off parameter of $l_{cut}=11$ \eqref{eq:fdip}.}
        \label{fig:butterflys}
\end{figure*}
\begin{figure*}[h!]
       \centering        
 \begin{subfigure}[b]{0.32\textwidth}
         \centering
        \caption{}
         \includegraphics[width=\textwidth]{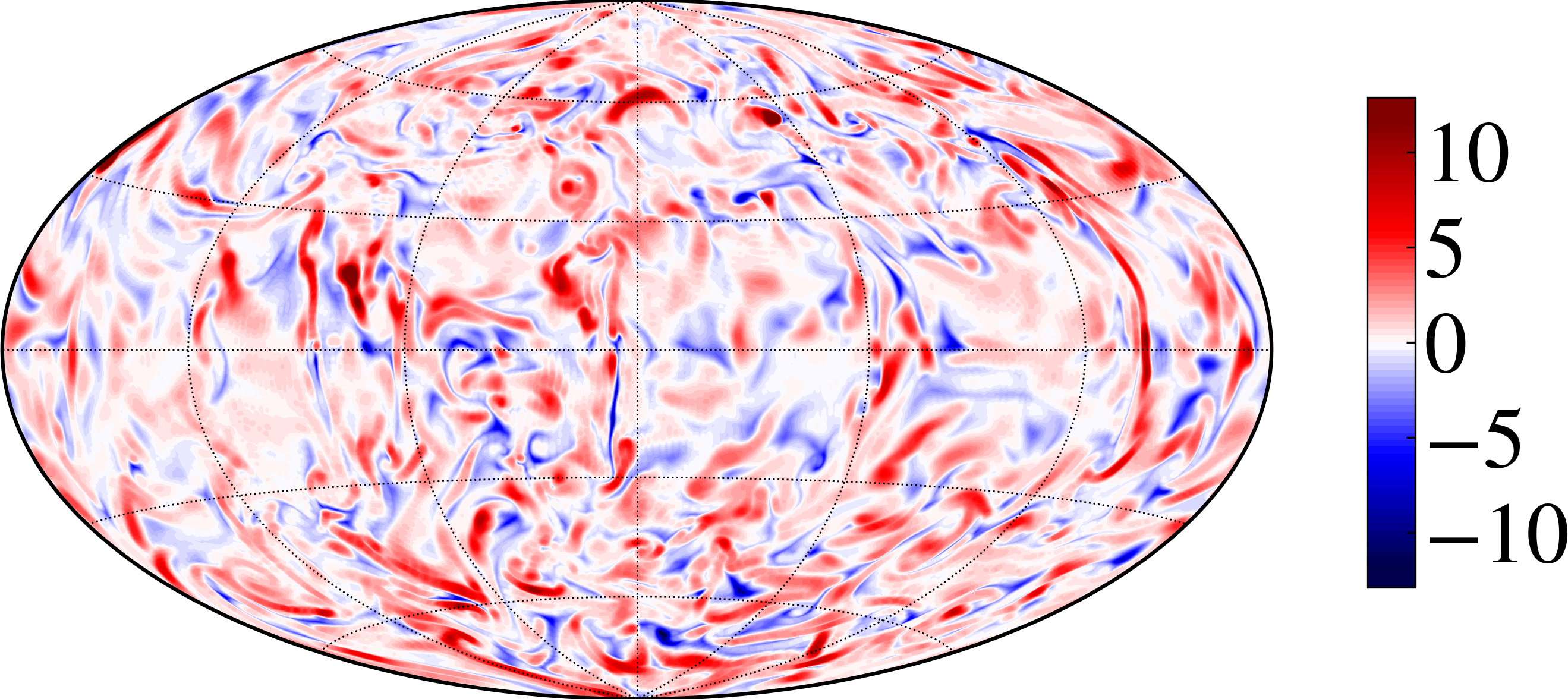}
     \end{subfigure}
 \hfill
     \begin{subfigure}[b]{0.32\textwidth}
         \centering
        \caption{}
         \includegraphics[width=\textwidth]{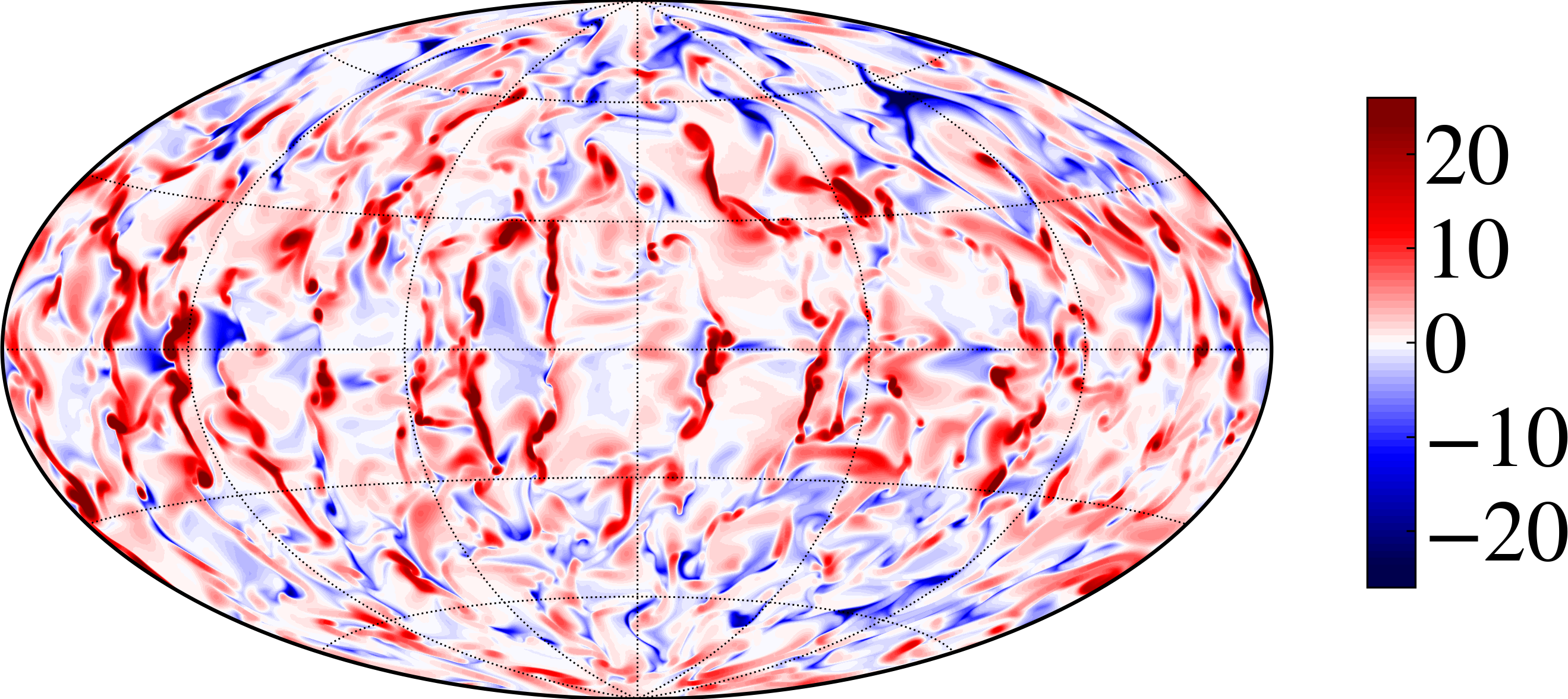}
     \end{subfigure}
      \hfill
     \begin{subfigure}[b]{0.32\textwidth}
         \centering
        \caption{}
         \includegraphics[width=\textwidth]{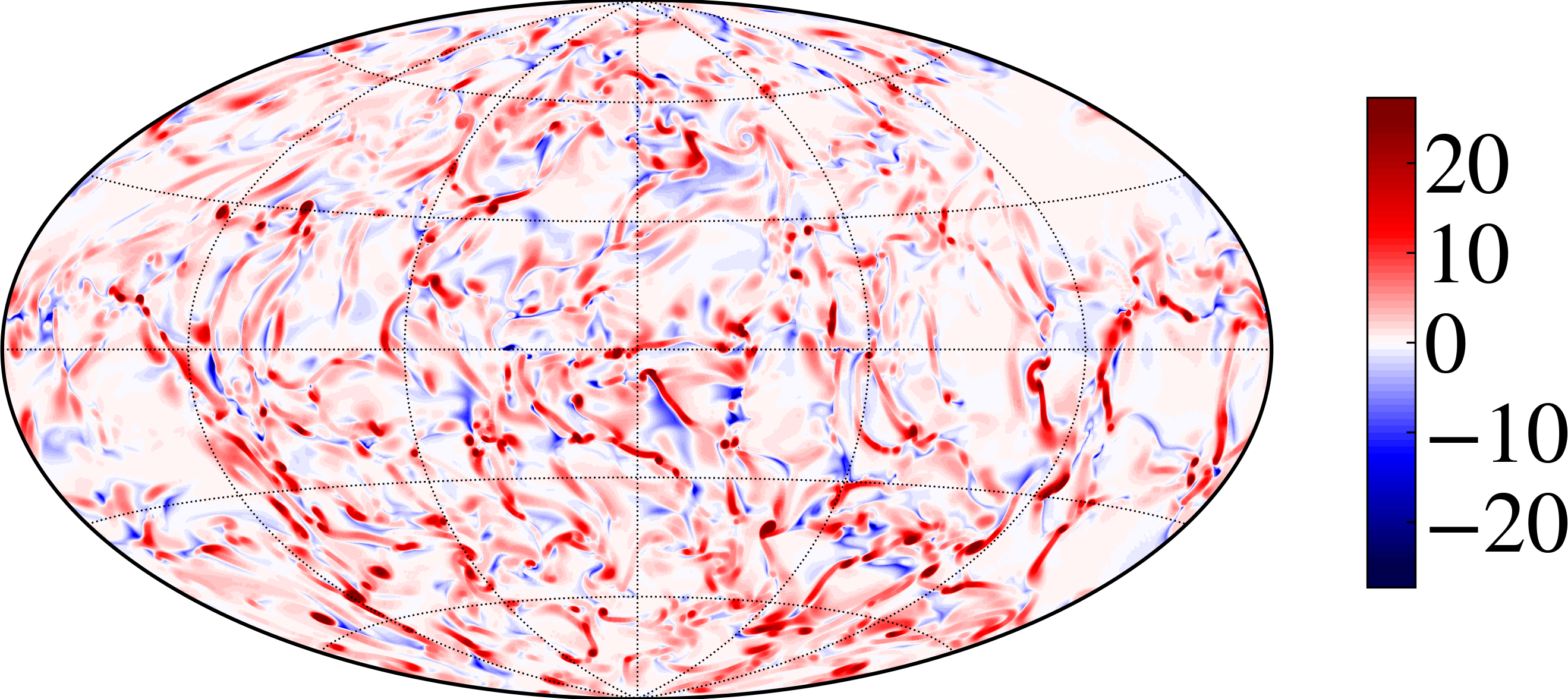}
     \end{subfigure}
     \vfill 
\begin{subfigure}[b]{0.32\textwidth}
         \centering
        \caption{}
         \includegraphics[width=\textwidth]{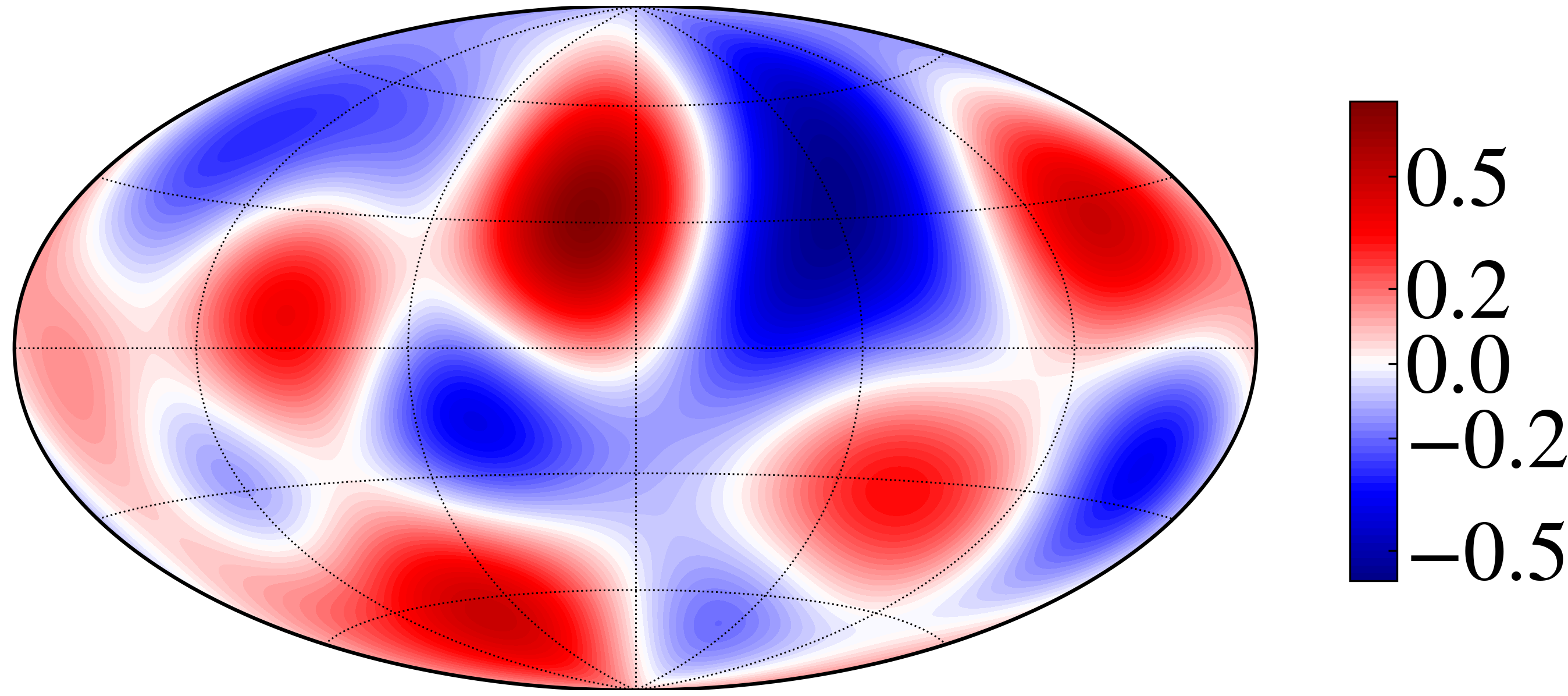}
     \end{subfigure}
 \hfill
 \begin{subfigure}[b]{0.32\textwidth}
         \centering
        \caption{}
         \includegraphics[width=\textwidth]{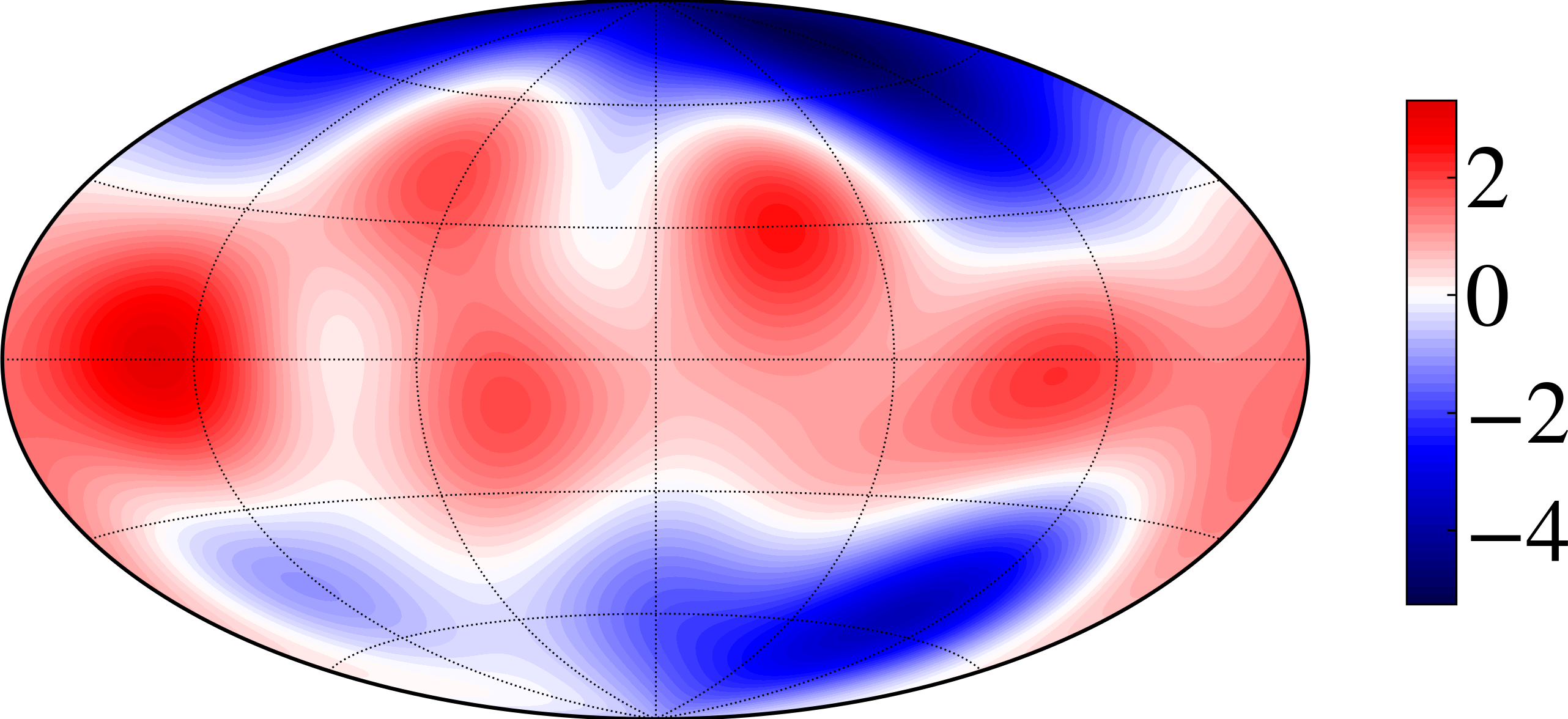}
     \end{subfigure}
     \hfill
   \begin{subfigure}[b]{0.32\textwidth}
         \centering
        \caption{}
         \includegraphics[width=\textwidth]{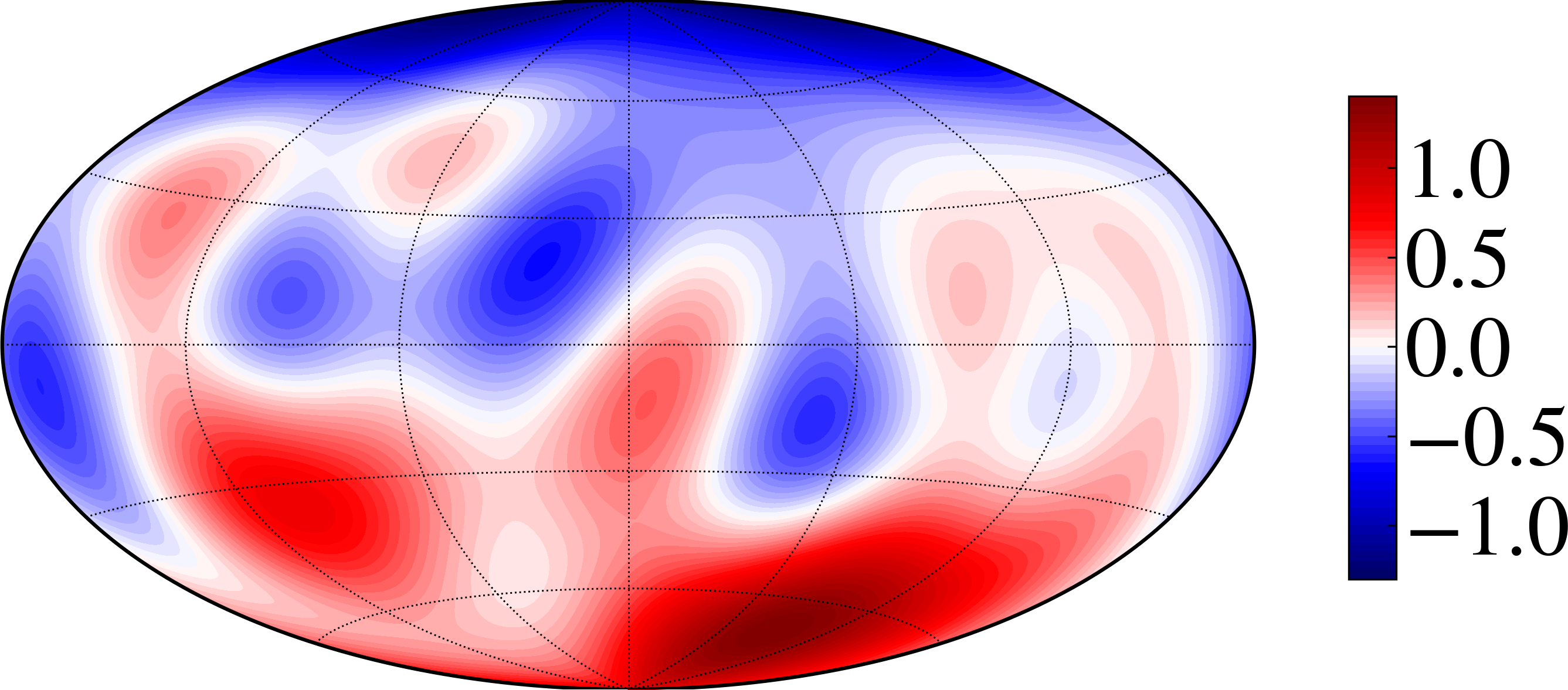}
     \end{subfigure}
 \caption{Instantaneous snapshots of the surface radial magnetic field $B_r$ (top), with the same fields filtered to $l\leq 5$ (bottom). (a,d) $N_\rho = 6$, $Ra/Ra_{cr}=32$, multipolar fields; (b,e) $N_\rho = 6$, $Ra/Ra_{cr}=16$, quadupolar waves; (c,f) $N_\rho = 6$, $Ra/Ra_{cr}=32$, during dipole period. $r/r_o = 0.95$. Color maps in panels (a-c) were saturated to $0.5 |B_r|^{max}$ to highlight the spatial structure of magnetic features.}
\label{fig:filteredBr}
\end{figure*}

 \begin{figure*}
\centering
           \begin{subfigure}[b]{0.24\textwidth}
         \centering
        \caption{}
         \includegraphics[width=\textwidth]{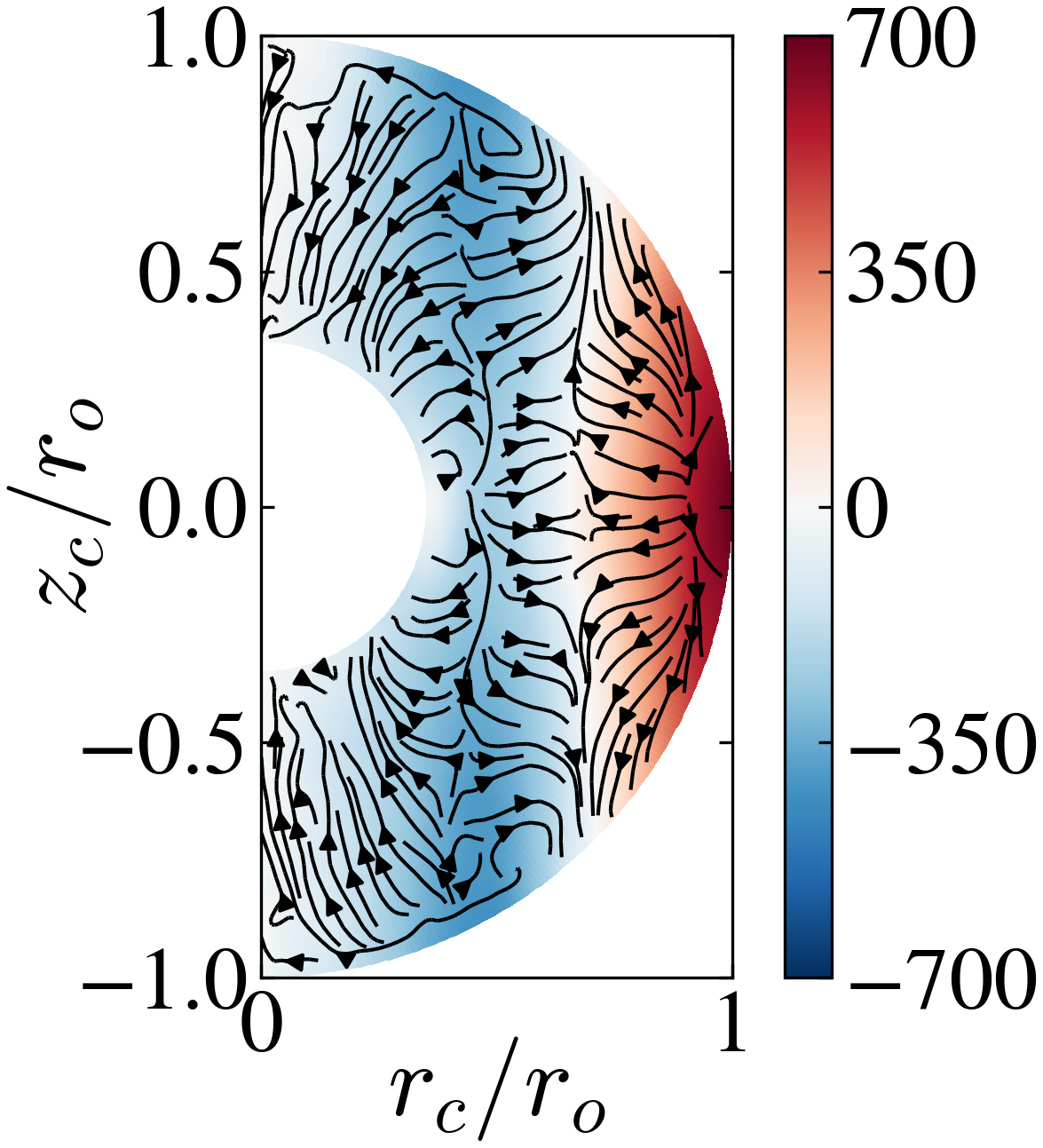}
           \end{subfigure}
         \hfill
     \begin{subfigure}[b]{0.245\textwidth}
         \centering
        \caption{}
         \includegraphics[width=\textwidth]{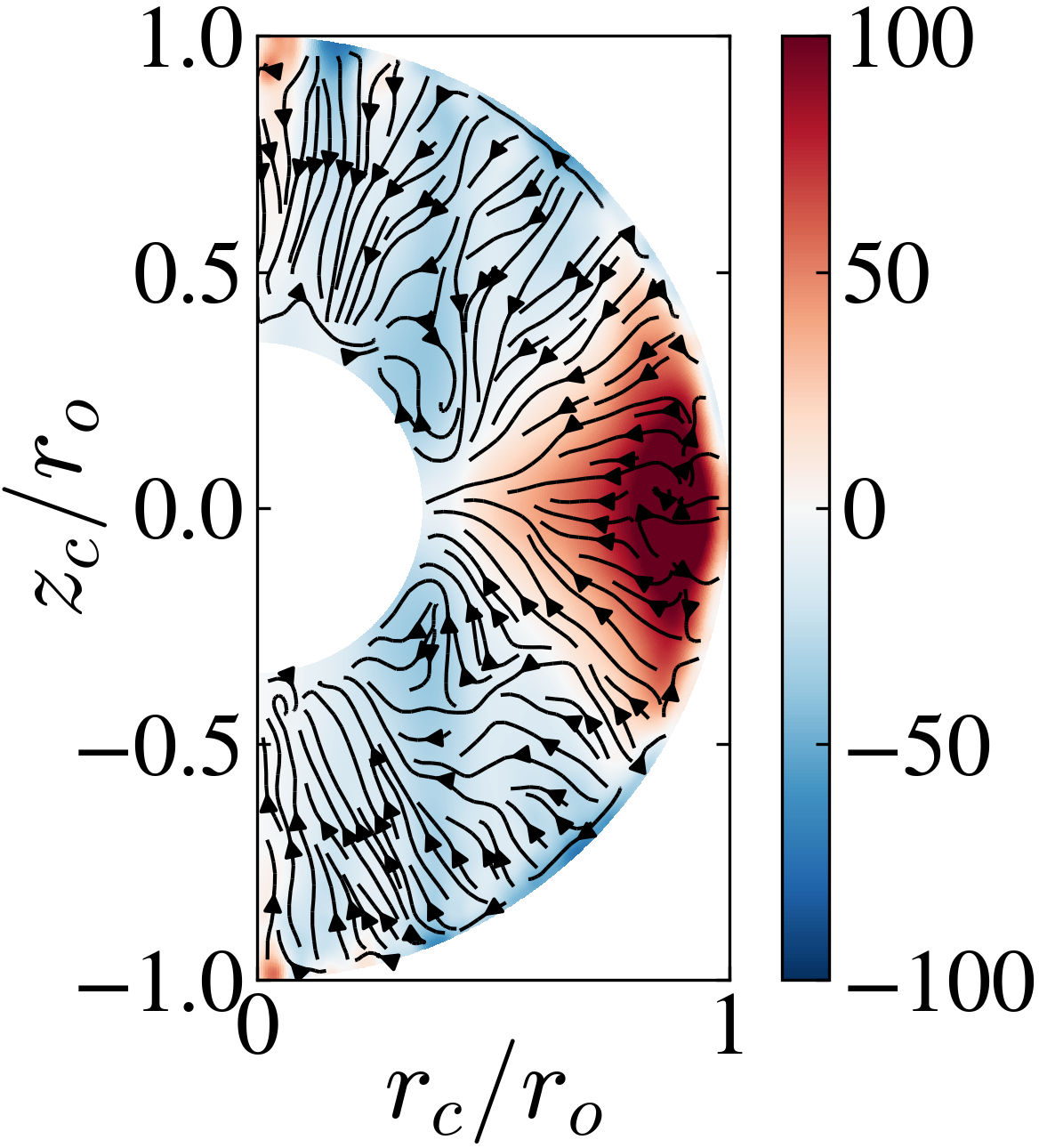}
        \end{subfigure}
           \hfill
               \begin{subfigure}[b]{0.24\textwidth}
         \centering
        \caption{}
         \includegraphics[width=\textwidth]{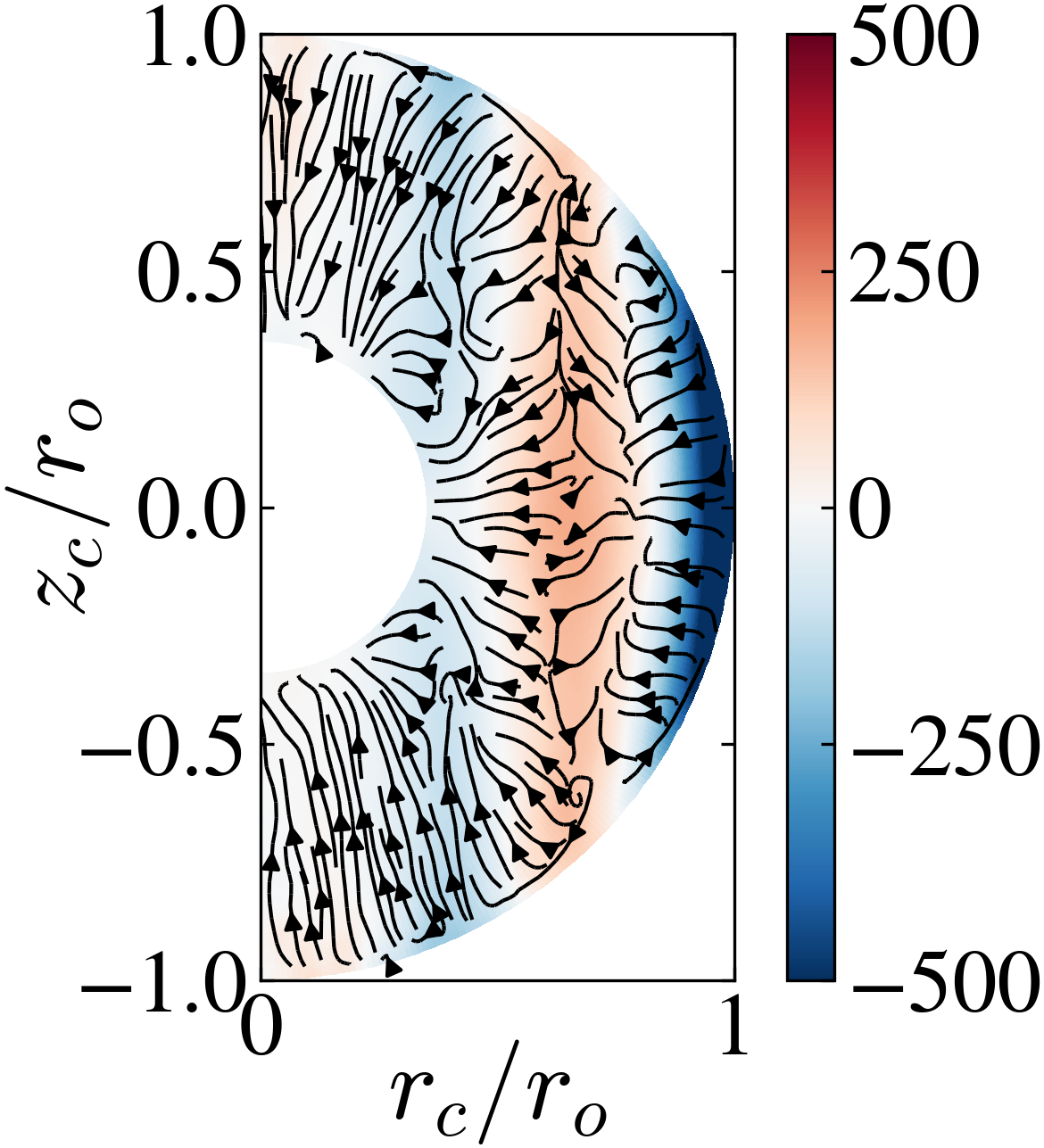}
     \end{subfigure}
     \hfill
     \begin{subfigure}[b]{0.245\textwidth}
         \centering
        \caption{}
         \includegraphics[width=\textwidth]{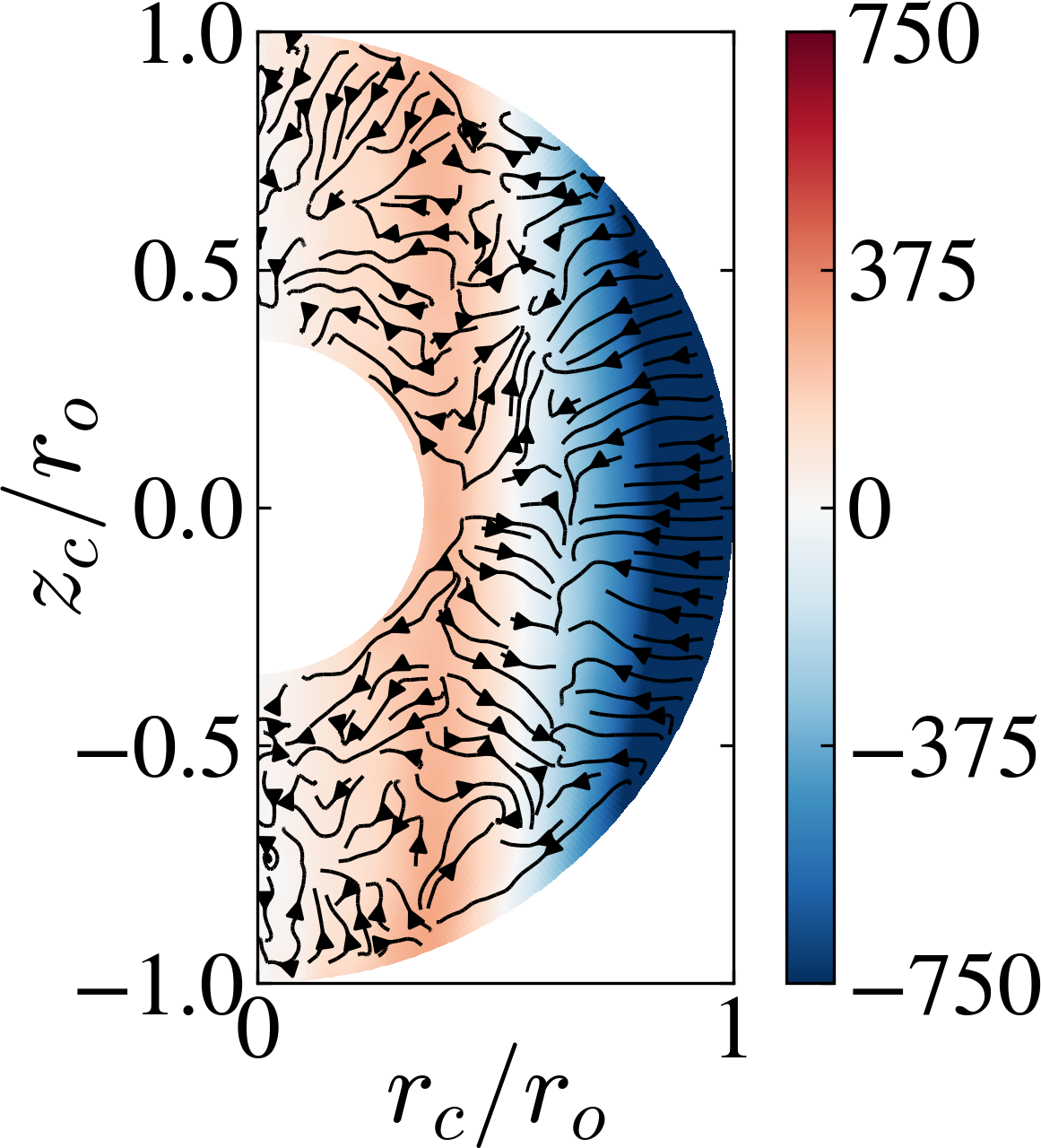}
     \end{subfigure}
        \caption{Zonal flows (in color)  for different types of dynamo solutions corresponding to figure~\ref{fig:butterflys}.  (a) $N_\rho=2$, $\widetilde{Ra}=32$; (b)  $N_\rho=6$, $\widetilde{Ra}=16$; (c) $N_\rho=6$, $\widetilde{Ra}=32$, during low magnetic energy interval with predominance of dipole on large scales; (d) $N_\rho=6$, $\widetilde{Ra}=32$, during high-energy interval without dipole. Flux lines corresponds to the poloidal component of Poynting flux (see section~\ref{sec:shear2flux}). Coordinates are denoted  in cylindrical coordinate system $(r_c,z_c)$. Fluxes and zonal flows were averaged in time over multiple snapshots.}
        \label{fig:flux_streamlines}
\end{figure*}
\subsection{Diagnostic quantities}
Butterfly diagrams in figure~\ref{fig:butterflys} present the dependence of the azimuthally averaged radial component of the magnetic field $\langle B_r \rangle_\phi$ at the surface of the outer sphere, $r=r_o$, as a function of rotational time $\Omega t$ and latitude $\theta$. They serve as a proxy for both  magnetic topology - dipolar or multipolar, and dynamics. A more quantitative representation of the dynamo topology is the surface dipolarity, 
\anna{
\begin{equation}\label{eq:fdip}
    f_{\rm dip}^{\ell_{\rm cut}} =\left(\dfrac{\sum_{m=-1}^{1} E^{l=1,m}_{mag} (r_o)}{\sum_{\ell=1}^{\ell_{\rm cut}} \sum_{m=-\ell}^{m=+\ell} E^{l,m}_{mag} (r_o)}\right)^{1/2} \; .
\end{equation}
}
which is the square root of the ratio between the energy contained in the dipolar spherical harmonic with $m\in [-1, 0, 1]$, $l=1$, and the energy contained in the first $l_{cut}$ spherical harmonics. The classic definition of a dipolar dynamo is $f_{dip}>0.5$; if $f_{dip}<0.5$, then the dynamo is considered multi-polar. We use here $l_{cut}=11$ as a default output of MAGIC code; we found that decreasing the cut-off length scale in \eqref{eq:fdip} to $l_{cut}=7$, perhaps more realistic for spectropolarimetric observations and ZDI, does not strongly affect the estimation of $f_{dip}^{l_{cut}}$. Together with $\langle B_r\rangle_\phi$ at the surface, figure~\ref{fig:butterflys} shows this quantity for different solution types.  The colormaps were saturated to highlight large scale fields; the maximum values of magnetic fluctuations close to the surface are actually much stronger in all our simulations. 

Spectropolarimetric observations typically allow to reproduce magnetic field topology up to $5-7$ spherical harmonics reliably, due to mutual cancellation of small-scale magnetic features \citep{kochukhov2021magnetic}. To provide a qualitative comparison of our models to these observations, we perform spectral filtering of surface radial magnetic field $B_r$, with $l_{cut} =5$ as a cut-off parameter.  Figure~\ref{fig:filteredBr} compares these large-scale filtered fields with the corresponding instantaneous maps of the radial magnetic field near the surface, $r/r_o = 0.95$.  

Finally, we identify the structure of differential rotation by the axisymmetric component of the toroidal velocity field $\langle u_\phi \rangle_\phi$, with brackets $\langle \cdots \rangle_\phi$ denoting a spatial average of $\phi$ direction. In the following, we refer to this quantity as zonal flows. The local sign of the zonal flow is directly related to the mean deviation $\langle u_\phi \rangle_\phi / r_c$ of the flow from the solid-body rotation $\Omega$. The zonal flows for different dynamo solutions in our simulations are presented in figure~\ref{fig:flux_streamlines}, where the corresponding fields were also averaged in time.

\subsection{Oscillatory dynamos in weakly stratified systems}\label{sec:multi_sol}

 The first type of dynamo behavior develops at low stratification, $N_\rho=2$, and corresponds to the less rotationally constrained runs with higher values of $Ro_l$ throughout the domain (green triangles in figure~\ref{fig:sim_param}b). It is a predominantly multipolar solution with a signature of distorted waves propagating toward the poles (figure~\ref{fig:butterflys}a). This waves, however, are very weak, compared to overall turbulent background (see Appendix~\ref{sec:app_sim_param} and table~\ref{tab:sim_param} therein).  In this regime, the surface dipolarity parameter remains always smaller than $0.5$, so this flow state lacks any prolonged dipole periods. Consistent with this, the instantaneous snapshots of $B_r$ reveal very small-scale fields from which it is hard to predict the distribution of their large scales (figure~\ref{fig:filteredBr}a). When filtered, it results in a  multi-polar configuration (figure~\ref{fig:filteredBr}d) on large scales, corresponding to low stratification. Such periodic dynamo solutions have been extensively studied in weakly stratified dynamo models, examples including \cite{schrinner2011oscillatory,schrinner2012dipole,karak2015magnetically,strugarek2017reconciling,viviani2019stellar}. They are usually attributed to Parker-Yoshimura dynamo waves driven by the mean-field $\alpha^2 \Omega$ or $\alpha \Omega$ dynamo mechanisms; the direction of their propagation is defined by the gradient of angular velocity. For this reason, here we do not focus on the particular properties of these waves.  See  the excellent reviews by \cite{brun2017magnetism} and \cite{charbonneau2020dynamo}, and the references therein for more details. 

 The differential rotation is solar-like at low $\widetilde{Ra}$ in this weakly-stratified case. Figure~\ref{fig:flux_streamlines}a shows the resulting profile of $\langle u_\phi \rangle_\phi$, averaged in time, for $\widetilde{Ra}=32$ (in color). Two opposite mean-flow directions develop, a prograde jet at the equator and a retrograde jet in the bulk. These flow structures are aligned vertically, and their relative strengths are comparable. Even though the flow transitions to anti-solar rotation for $\widetilde{Ra} \geq 80$, with inversed meridional distribution (see the inset on the right in figure~\ref{fig:sim_param}b), the average dipolarity of the flow does not increase with $\widetilde{Ra}$ (figure~\ref{fig:dipolarity}a) and the dynamo configuration remains non-dipolar.

\subsection{Oscillatory dynamos in rapidly rotating and strongly stratified systems (large $N_\rho$ and $Ro_l<1$ for all radii).}\label{sec:waves_sol}
 The second type of the dynamo solutions develops when stratification is increased to $N_\rho=4$ or $6$, but the levels of turbulence are weaker, e.g. $\widetilde{Ra}\leq 16$ (see the corresponding values of Reynolds number in table~\ref{tab:sim_param}, increasing with the level of supercriticality). These are rather strongly rotationally constrained simulations with $Ro_l<1$ throughout the domain (figure~\ref{fig:sim_param}a), corresponding to blue circles in figure~\ref{fig:sim_param}b. The dynamo is dominated in this case by  large-scale, clearly manifested dynamo waves (figure~\ref{fig:butterflys}b). These waves are predominantly symmetric with respect to equator; in the following, we will refer to such symmetry of magnetic field as quadrupolar. The time scale of these waves is much slower than in previous, weakly-stratified case ($N_\rho=2$), with about $T/T_{rot}=500$ for wave period as opposed to at most $T/T_{rot}=100$ for weakly stratified flows. Compared to the dynamo configuration in figure~\ref{fig:butterflys}a, the waves are much stronger and so their spatiotemporal coherency is visible from magnetic fluctuations at the surface~\ref{fig:filteredBr}(b). When filtered from small-scales, large-scale magnetic field in this case represent a quadrupolar, axisymmetric field (figure~\ref{fig:filteredBr}e), with surface distribution reflecting the moment in time with respect to the total period of the wave (figure~\ref{fig:butterflys}e).  Contrary to predominantly multipolar solution in figure~\ref{fig:butterflys}a, in the quadrupolar waves state all the large-scale modes exhibit nearly periodic behavior, as is reflected in the oscillation of dipolar component. The dipolarity in this case is also quite small and does not exceed $f_{dip}^{11} \leq 0.3$ during its maxima. 

Such oscillatory dynamos also exhibit solar-like differential rotation (figure~\ref{fig:flux_streamlines}b), although the topology of the zonal flow is different. In this case, the prograde jet has a conical shape and is confined predominantly to the area close to the equator; it exhibits periodic oscillations, i.e. expands and shrinks during magnetic cycle in figure~\ref{fig:butterflys}b. The opposing, retrograde flow in the bulk is much weaker than in the previous, low-stratified case; the overall magnitude of the zonal flow also remains weak.

\subsection{Temporal variations of the magnetic field symmetry in strongly stratified and turbulent systems}\label{sec:dip_sol}

Finally, we describe the third type of dynamo solutions that develops at high levels of turbulence, $\widetilde{Ra}=32$ and high stratification, $N_\rho=4$ or $6$, featuring surface layers with $Ro_l>1$ (figure~\ref{fig:sim_param}a). These simulations are thus the most representative of stellar flow regimes, especially for slowly rotating stars, and corresponds to the diamond symbols in figure~\ref{fig:sim_param}b. It represents an aperiodic dynamo with time intervals of predominantly dipolar configuration (i.e. anti-symmetric with respect to the equator) and intervals of multipolar magnetic solutions without visible domination of any particular spherical harmonic (figure~\ref{fig:butterflys}c). During strong dipolar events, the surface dipolarity is large and reaches the values of $f_{dip}^{11} = 0.5-0.6$, formally corresponding to dipolar magnetic fields. The dipolar component is the most intense at high latitudes, and less visible near the equator, where magnetic behavior is less coherent. 

\begin{figure*}
    \centering
    
    \includegraphics[width=\linewidth]{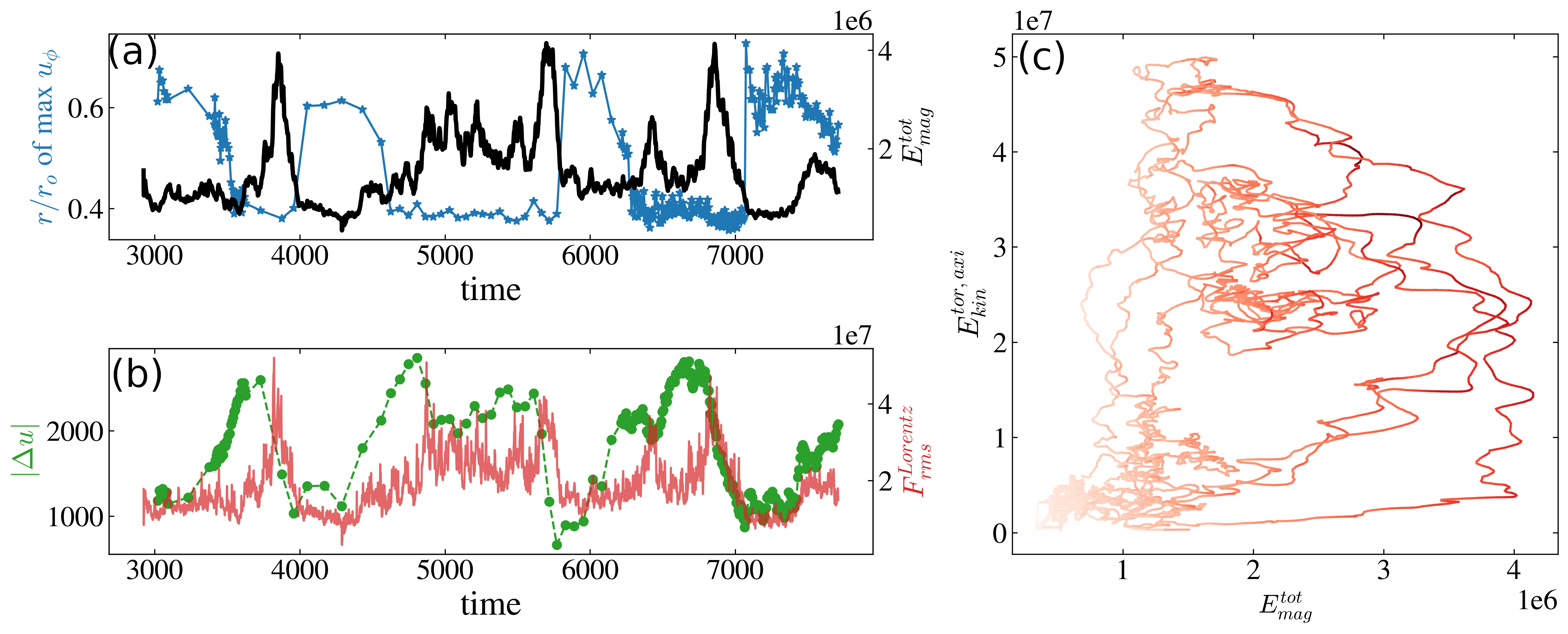}
    \caption{(a) Total magnetic energy as a function of time (in black) vs the radius of $\langle u_\phi \rangle_\phi^{max}$ in the equatorial plane ($\theta = \pi/2$), in blue. Large radii correspond to 3-layer rotation, while low values denote anti-solar rotation (compare figures~\ref{fig:flux_streamlines}c,d). (b) The magnitude of the zonal flow, $\Delta u^{eq}_\phi = |\langle u_\phi \rangle^{max} - \langle u_\phi \rangle^{min}|$, in the equatorial plane (in green). Rms of the magnetic Lorentz force, in red. (c) The two energy states of the flow - high and low - in the phase space of total magnetic energy and the energy of axisymmetric (zonal) flow. The points are colored by the rms of the Lorentz force. $N_\rho=6$, $\widetilde{Ra}=32$.}
    \label{fig:phase_space}
\end{figure*}
The large-scale, dipole component exhibits active dynamics, with two subsequent magnetic reversals around $\Omega t = 3600$, then loss of coherent magnetic field and development of the multipolar magnetic state between $\Omega t \in [4500, 5500]$, a subsequent dipole re-appearance at about $\Omega t=5500$, reversal, and loss at $\Omega t=6000$. We ran these simulations for at least one-two magnetic diffusion time scale, and observed that this dynamics  persists in time. Similar behavior takes place at lower density stratification, $N_\rho=4$, and strong convection $\widetilde{Ra}=32$.

When filtered, magnetic fields show predominantly dipolar configuration at $N_\rho=6$, $\widetilde{Ra}=32$ during a dipolar interval (figure~\ref{fig:filteredBr}f). During non-coherent, multipolar intervals, or dipole reversals, the field topology appears similar to the one for $N_\rho=2$ in panel (d). Compared to the total field magnitude in the upper panels, unfiltered fields are order of magnitude stronger for all the three cases.

In this case,  two different types of differential rotation are observed at different times in the same simulations. During dipolar periods, a thin retrograde jet develops at the surface, and the prograde jet shifts to the mid-gap between the spheres. The inner flow area, including the area inside the tangent cylinder, develops a very weak retrograde rotation and remains quiescent  overall (figure~\ref{fig:flux_streamlines}c). In the following, we will refer to this configuration as three-layer differential rotation. During multipolar periods, the dynamo entirely transitions to stronger anti-solar differential rotation, with a retrograde jet expanding at the surface and prograde jet shifting into the bulk of the domain, toward the tangent cylinder (figure~\ref{fig:flux_streamlines}d). The peak of the shear at the boundary between the surface retrograde jet and inner prograde jet thus also shifts in the interior of the domain.

We note here that these aperiodic dipolar solutions exhibit bi-stability with respect to periodic waves from section~\ref{sec:waves_sol} at $N_\rho=6$, $\widetilde{Ra}=32$, as shown in figure~\ref{fig:butterflys}(b), depending on initial conditions. Slightly increasing $Ra$ would potentially eliminate this bi-stability.

\section{Dynamical properties of the flow and the field in strongly stratified and turbulent systems}\label{sec:cycle}

\subsection{Predator-prey interaction of zonal flows and magnetic  fields}\label{sec:cycle2uphi}
In this section, we focus our attention on the dynamics of strongly stratified and turbulent aperiodic dynamo states, described in the previous section~\ref{sec:dip_sol}. To monitor transitions between the two types of differential rotation in this run, three-layer and anti-solar, we follow the radial position of $u^{max}_\phi$ at the equator, $\theta = \pi/2$, i.e. the maximum in the prograde jet. Since both configurations of zonal flows are aligned with the rotation axis (figure~\ref{fig:flux_streamlines}c,d), $u^{max}_\phi$ is a good proxy for the type of the zonal flow: $r/r_o\sim 0.4$ for anti-solar and $r/r_o \sim 0.65$ for three-layer configuration. We plot this quantity in figure~\ref{fig:phase_space}a, together with the total magnetic energy of the flow, $E^{tot}_{mag}$. It appears that the total magnetic energy correlates well with the type of the zonal flow, developing strong peaks when rotation is anti-solar, and local minima otherwise.  Therefore, these dynamo regimes exhibit a bi-stability between the flow state with low magnetic energy and three-layer differential rotation, and the flow state with high magnetic energy and two-layer, anti-solar zonal flow.  The dynamo transitions between the two states aperiodically, spending in each state about $500-1000$ rotation times. However, these runs are computationally too expensive to gather accurate temporal statistics for these stochastically driven transitions.

To explain the physical mechanism behind these reversals, i.e. whether magnetic fields are driving re-configuration of the zonal flows, we clarify the causality between the growth and decay of the magnetic field and the amplitude of the zonal flows. To this end, we plot a proxy for the strength of zonal flows,  $\Delta u^{eq}_\phi = |\langle u_\phi \rangle^{max} - \langle u_\phi \rangle^{min}|$ at the equator, together with the rms of the Lorentz force $\mathbf{F}^{Lorentz} \propto (\nabla \times \mathbf{B}) \times \mathbf{B}$, as a function of time in figure~\ref{fig:phase_space}b. The strength of the shear increases by a factor of two during the transitions from three-layer to anti-solar rotation, well before the peaks in $E_{mag}^{tot}$ develop. This indicates that the development of anti-solar rotation that temporally enhances magnetic energy. On the other hand, it is when magnetic field is the strongest that the magnitude of differential rotation begins to decrease; these peaks coincide with the maxima in the rms of the Lorentz force, acting thus as magnetic damping on the zonal flow. Subsequently, the system returns to low-energy, three-layer rotation state. Figure~\ref{fig:phase_space}c summarizes these dynamical transitions in the phase space of the total magnetic energy, the axisymmetric kinetic energy of the zonal flow, and the Lorentz force.

\begin{figure*}
    \centering
 \begin{subfigure}[b]{0.3\textwidth}
         \centering
        \caption{}
         \includegraphics[width=\textwidth]{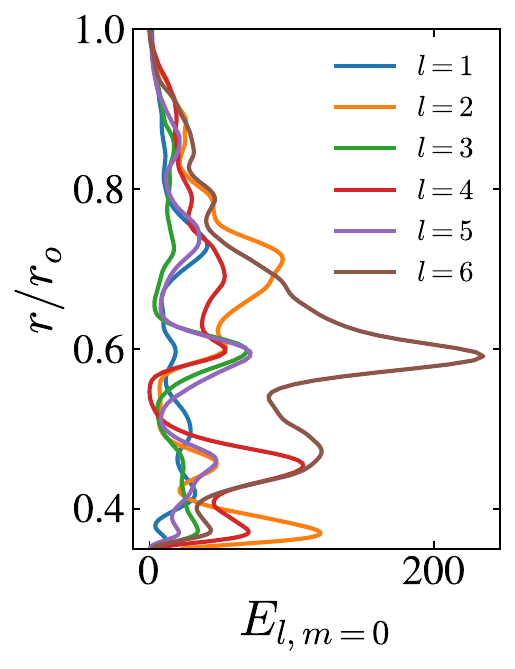}
     \end{subfigure}
\hfill 
      \begin{subfigure}[b]{0.3\textwidth}
         \centering
        \caption{}
         \includegraphics[width=\textwidth]{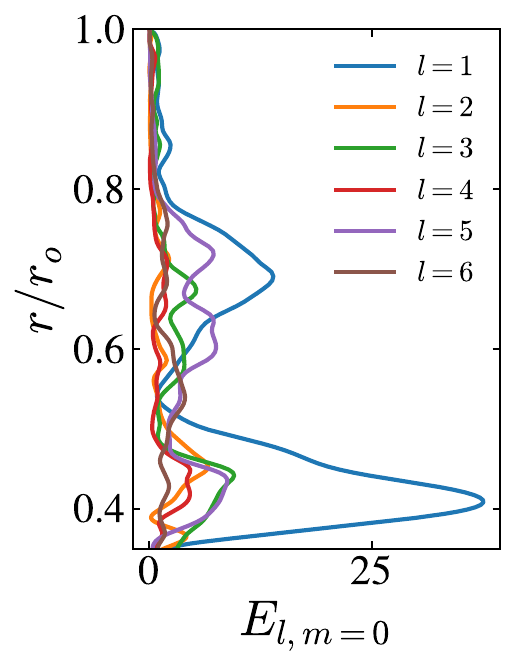}
     \end{subfigure}
\hfill 
 \begin{subfigure}[b]{0.3\textwidth}
         \centering
        \caption{}
         \includegraphics[width=\textwidth]{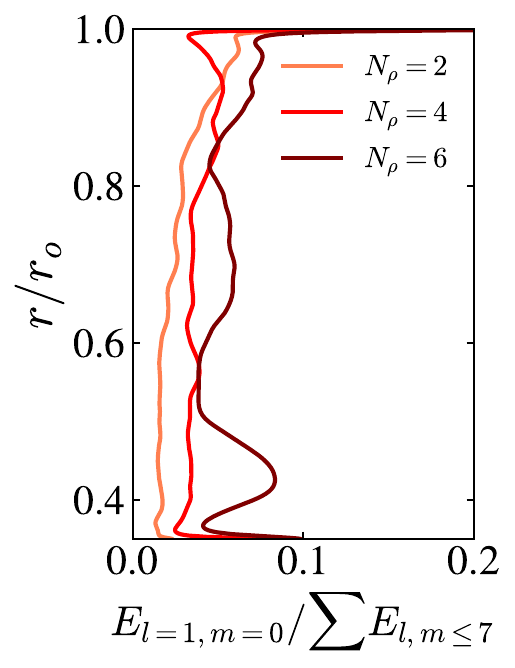}
     \end{subfigure}
    \caption{ (a,b) \anna{Magnetic} energy in the first six axisymmetric spherical harmonics ($m=0$), as a function of spherical radius $r$, at $N_\rho=6$, $\widetilde{Ra}=32$. (a) High-energy state, anti-solar rotation, $t_{rot} \approx 6900$.  (b) Low-energy state, 3-layer rotation,  $t_{rot}\approx 7300$. The relative build-up in dipolarity with respect to large scales takes place only during low-energy states with weakened 3-layer rotation. (c) The energy in axisymmetric dipole normalized with the total large-scale magnetic energy up to 7th spherical harmonic, $\sum E_{l,m\leq7}$, at $\widetilde{Ra}=32$. The modal energy was averaged over dipolar intervals for $N_\rho=4,6$, and over the whole time evolution for $N_\rho=2$. }
    \label{fig:int_dipolarity}
\end{figure*}

\begin{figure*}[h!]
\centering
    \begin{subfigure}[b]{0.45\textwidth}
         \centering
        \caption{}
         \includegraphics[width=\textwidth]{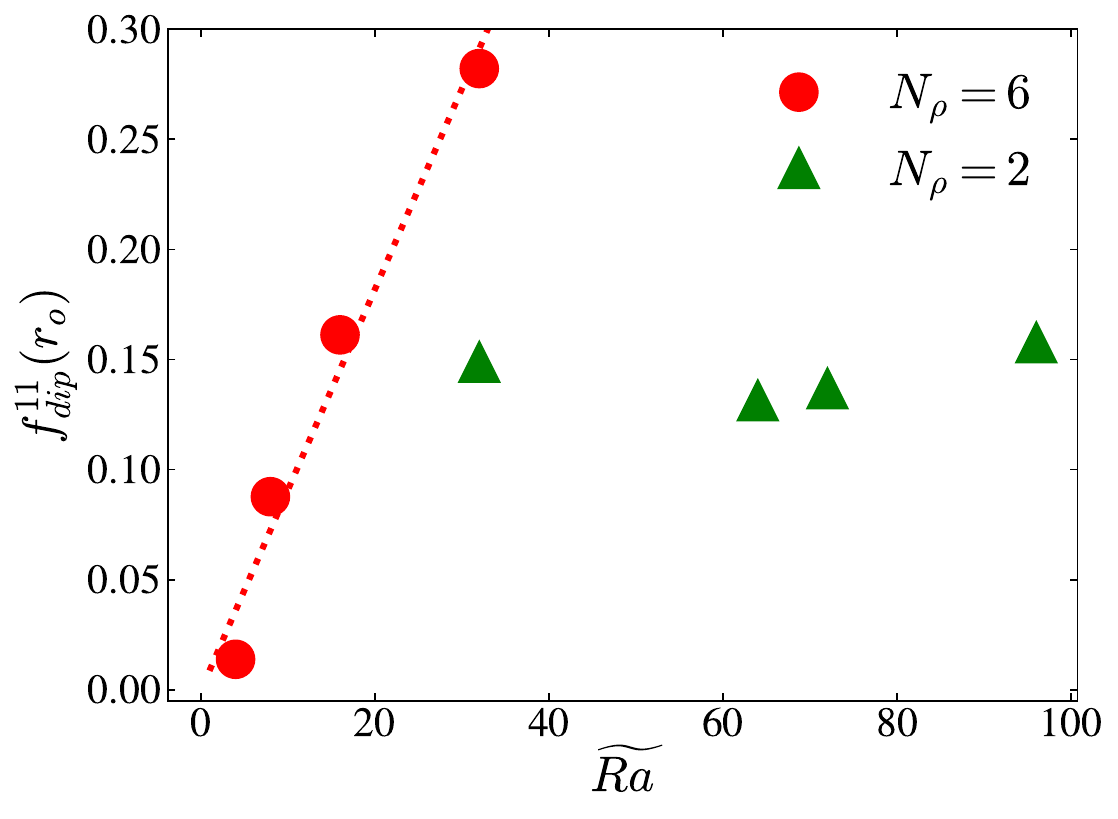}
     \end{subfigure}
     \hfill
         \begin{subfigure}[b]{0.45\textwidth}
         \centering
        \caption{}
         \includegraphics[width=\textwidth]{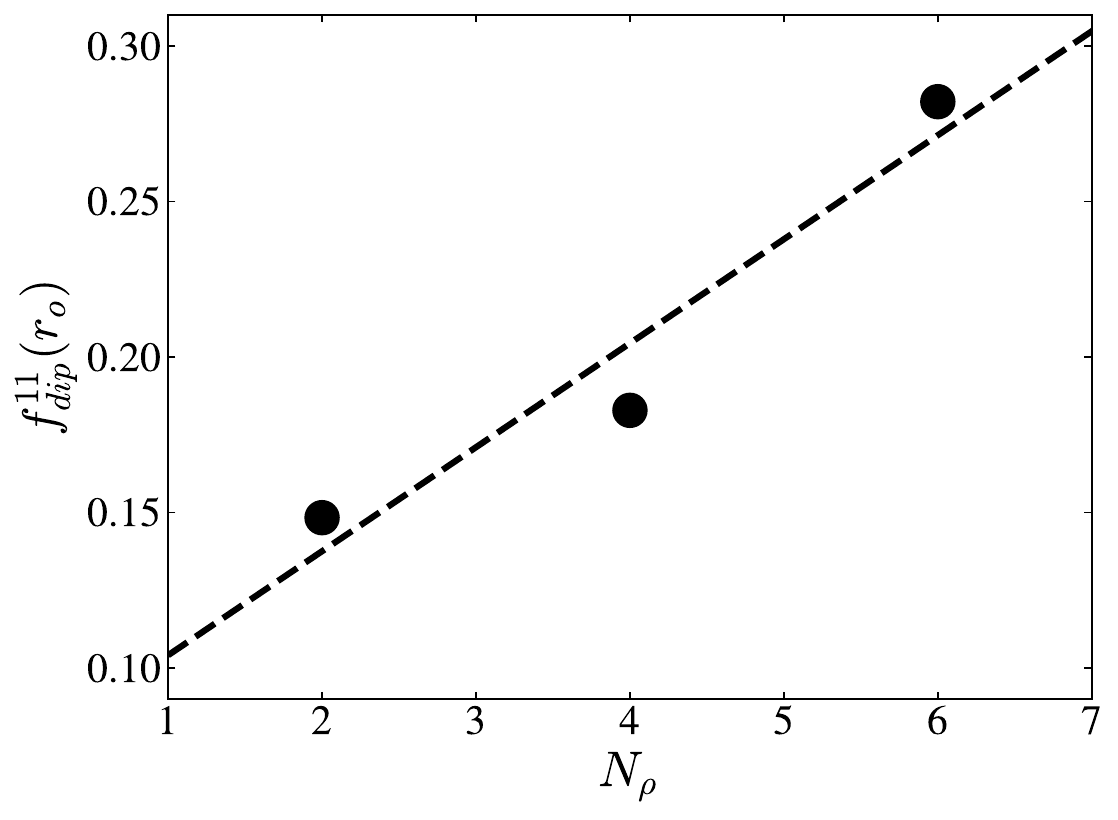}
     \end{subfigure}
         \caption{Averaged in time surface dipolarity of the dynamo solutions in figure~\ref{fig:butterflys}, as given by equation~\eqref{eq:fdip}. (a) As a function of $Ra$ at $N_\rho=6$ (circles) and $N_\rho=2$ (triangles); (b) As a function of $N_\rho$ at $Ra=32$. The averages are performed over dipolarity time series like those in figure~\ref{fig:butterflys}.}
        \label{fig:dipolarity}
\end{figure*}

\subsection{Correlation between the flow state and strength of dipolar modes}\label{sec:cycle2dip}

Finally, we link the flow cycles in section~\ref{sec:cycle2uphi} and the appearance of large-scale axisymmetric magnetic dipoles in section~\ref{sec:dip_sol}. At first, we compare the radial distribution \anna{of magnetic energy contained in} the first $6$ spherical harmonics with azimuthal wavenumber $m=0$, corresponding to the large-scale axisymmetric \anna{field} structures, during high- and low-energy states for the run with $N_\rho=6$, $\widetilde{Ra}=32$.

Figure~\ref{fig:int_dipolarity}a shows the radial profile of these modal energies at a moment in time ($\Omega t \approx 6900$) during the high-energy state with strong anti-solar rotation. The energy in all large-scale harmonics are comparable, with relatively small-scale component of $l=6$ peaking in the middle of the computational domain. On the other hand, the energy of axisymmetric dipolar mode with $l=1$, $m=0$ is much larger than the rest during a later moment with low magnetic energy and suppressed, three-layer differential rotation, even if the energy content decays by a factor of ten in all modes (figure~\ref{fig:int_dipolarity}b). The energy of the dipolar mode peaks at small radii of $r/r_o \approx 0.4$ (figure~\ref{fig:int_dipolarity}b), which receive their energy contribution predominantly from the quiescent areas within the tangent cylinder (figure~\ref{fig:flux_streamlines}c). This indicates that axisymmetric magnetic flux accumulates within the tangent cylinder, consistently with stronger surface fields at high latitudes from the butterfly diagrams (figure~\ref{fig:butterflys}c).  The first minima of the dipolar mode correlates with the location of the prograde jet in figure~\ref{fig:flux_streamlines}c. When the axisymmetric modes in figure~\ref{fig:int_dipolarity}b are normalized by the radial distribution of the cumulative large-scale energy, the dipolar mode is still consistently larger then the rest across the domain.

There is also a systematic increase in axisymmetric dipole energy with increase of density stratification. To see it, we average the axisymmetric dipolar mode with $l=1$, $m=0$ over several snapshots during dipolar periods and normalize it by the total large-scale magnetic energy up to the 7th spherical harmonic, $ \sum E_{l,m \leq 7}$. The averages are taken over the low-energy dipole intervals for $N_\rho=4$ and $6$ with three-layer differential rotation, and over the whole time integration of $N_\rho=2$, since it does not exhibit intervals of predominant dipolarity.  In figure~\ref{fig:int_dipolarity}(c), we compare this relative radial distribution of axisymmetric dipoles for the three values of $N_\rho$. The strength of the dipolar component indeed consistently increases with density stratification across the whole domain, especially at low radii. The build-up of dipolarity at the surface is correlated with the increase of the dipolar energy in the bulk in the low-energy states, although it is partially conditioned by the magnetic boundary conditions - the energy of all magnetic modes decays at the surface  (see figure~\ref{fig:int_dipolarity}a,b). For this reason, we find that relative volume-distributed characteristics of modal energy, such as those in figure~\ref{fig:int_dipolarity}(a,b), are better indicators of  low- and high-energy dynamo states in figure~\ref{fig:phase_space} than, for example, local in time, weaker  peaks in surface dipolarity in figure~\ref{fig:butterflys}.

\section{Mechanisms of dipole field generation}\label{sec:mechanisms}

\subsection{Build-up of dipolarity with increase of $N_\rho$ and $\widetilde{Ra}$}\label{sec:dipolarity}

Even though only flow states like those in figure~\ref{fig:butterflys}(c) have prolonged  intervals of dipolar magnetic topology in our simulations, the average values of dipolarity parameter $f_{dip}^{11}$ systematically increases with $\widetilde{Ra}$ and $N_\rho$. To illustrate this, we plot in figure~\ref{fig:dipolarity} the values of $f_{dip}^{11}$, averaged in time, for the three different paths in the parameter space (figure~\ref{fig:sim_param}b). In the first one, with low stratification, $N_\rho=2$ and variable convective strength $\widetilde{Ra}$, the relative strength of dipolar mode remains approximately the same even when the convection is three times stronger (figure~\ref{fig:dipolarity}a). In the second one, with high stratification $N_\rho=6$ (same figure), dipolarity linearly increases with Rayleigh number with a scaling of $f_{dip}^{11} ~\sim \widetilde{Ra}^{0.01}$. The third path, where convection intensity is fixed, $\widetilde{Ra}=32$, and stratification is varied, is shown in figure~\ref{fig:dipolarity}b. It also exhibit a linear scaling of $f_{dip}^{11} \sim N_\rho^{0.03}$, with a somewhat stronger exponent than the scaling with $\widetilde{Ra}$. Note that these are the mean values of dipolarity; its instantaneous maxima during dipolar intervals are higher (figure~\ref{fig:butterflys}). With $f_{dip}^{11}$ at $N_\rho =6$ and $\widetilde{Ra}=32$ as a reference, these results suggest that the  strength of dipolar mode would systematically increase further with a simultaneous increase in both $N_\rho$ and $\widetilde{Ra}$, and thus for realistic stellar densities further away from convective onset.  In the rest of this section, we aim to explain the mechanisms of this gradual build-up in dipolarity.

\subsection{Surface compression of small-scale magnetic fields}\label{sec:compression}
\begin{figure*}[h!]
        \centering
            \begin{subfigure}[b]{0.45\textwidth}
         \centering
        \caption{}
         \includegraphics[width=\textwidth]{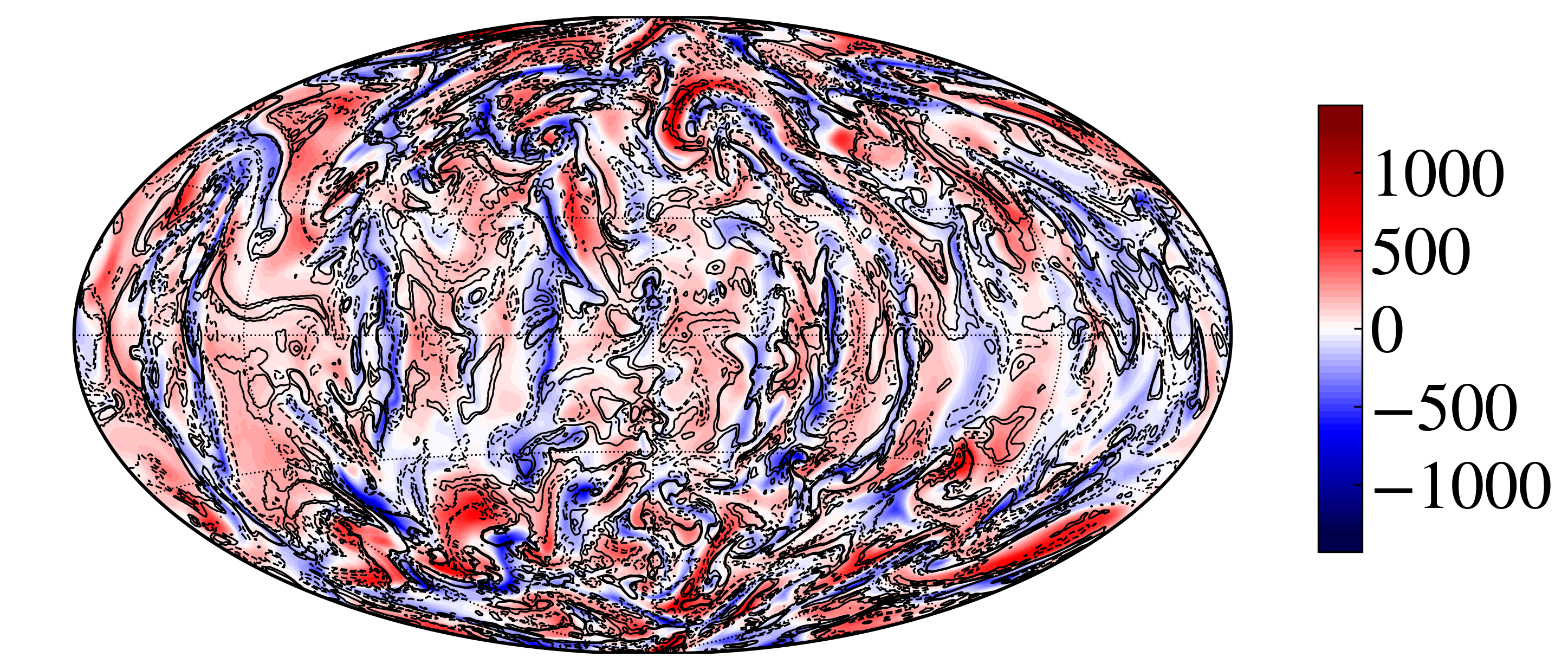}
     \end{subfigure}
     \hfill 
 \begin{subfigure}[b]{0.45\textwidth}
         \centering
        \caption{}
         \includegraphics[width=\textwidth]{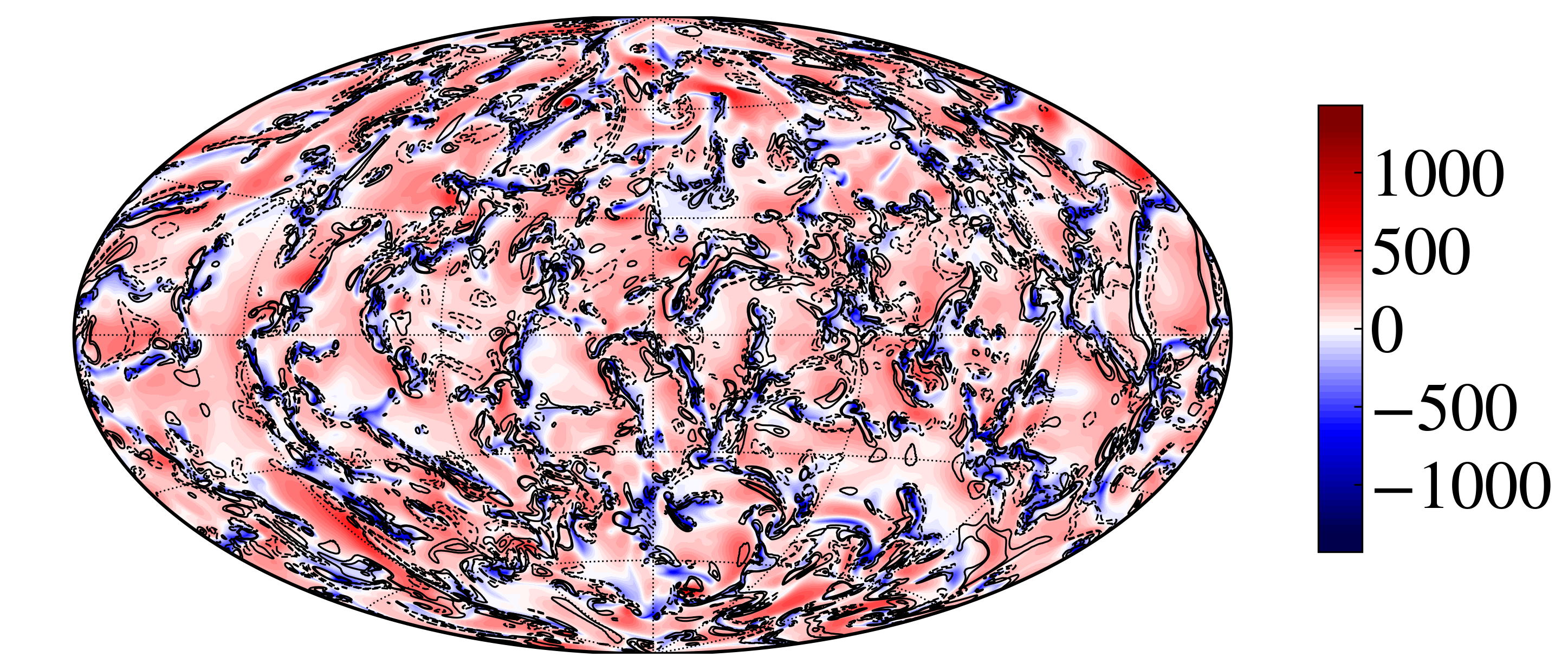}
     \end{subfigure}
    \vfill
          \begin{subfigure}[b]{0.45\textwidth}
    \centering
        \caption{}
         \includegraphics[width=\textwidth]{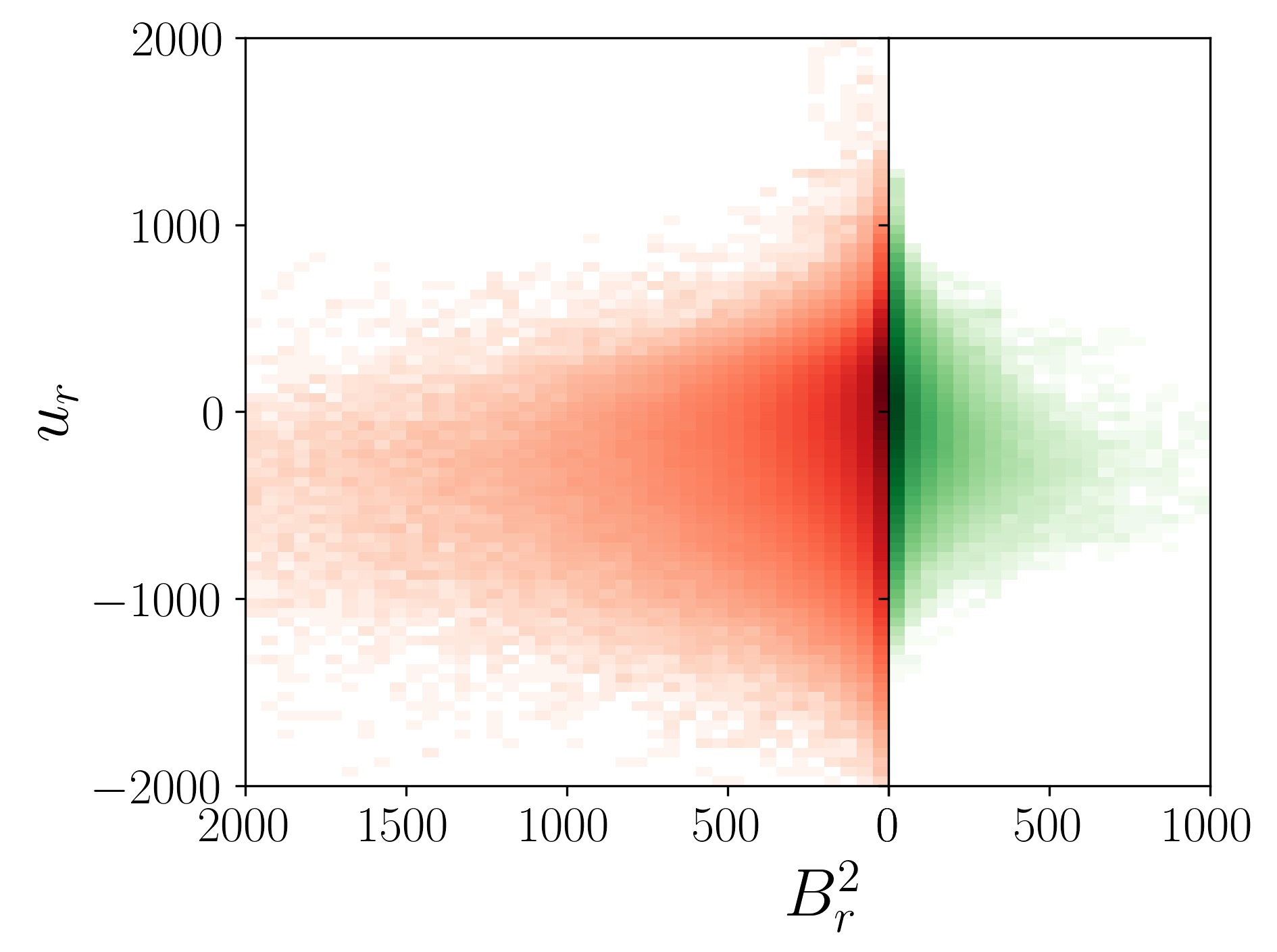}
     \end{subfigure}
     \hfill
    \begin{subfigure}[b]{0.45\textwidth}
    \centering
        \caption{}
         \includegraphics[width=\textwidth]{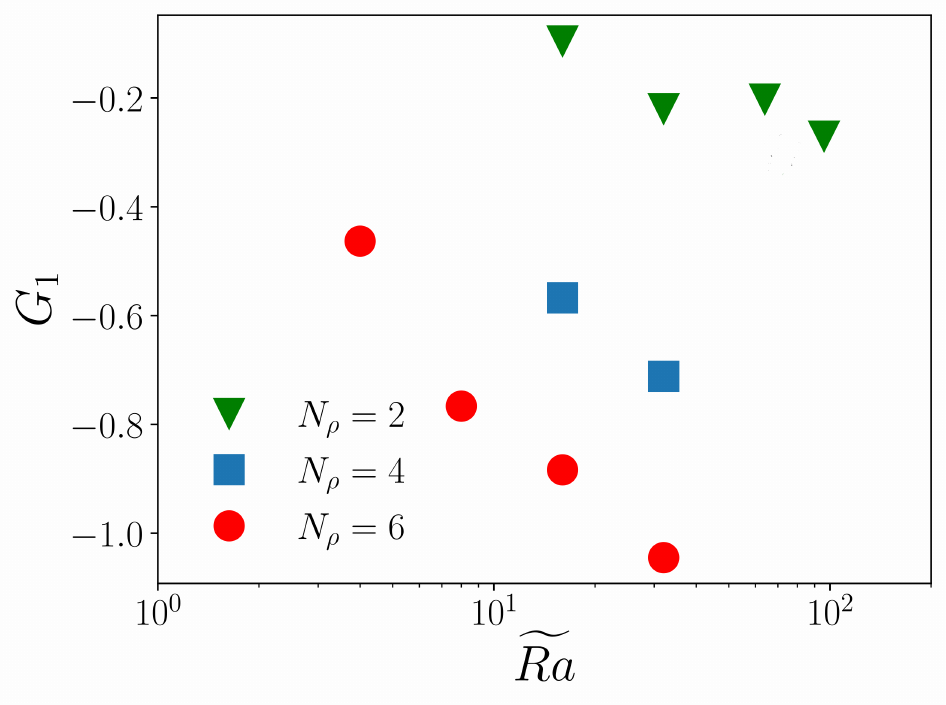}
     \end{subfigure}
             \caption{ (a) Instantaneous snapshots of radial magnetic field $B_r$ (contours) and radial velocity component $v_r$ (color) for $N_\rho=2$, $Ra/Ra_{cr}=32$ at $r/r_o = 0.9$. The contour levels are $[0.1 B_r^{min}, 0.05 B_r^{min}, 0.05 B_r^{max}, 0.1 B_r^{max}]$. (b) The same for stronger stratification, $Ra/Ra_{cr}=32$, $N_\rho=6$. (c) 2D probability density of the joint distribution of $b^2$ and $v_r$ near the surface, $r/r_o = 0.9$. Left, $N_\rho=6$; right, $N_\rho=2$. $\widetilde{Ra}=32$. Colors are spaced logarithmically. (d) Unbiased sample skewness coefficient $G_1$ \eqref{eq:skewness} as a function of $\widetilde{Ra}$. Red, $N_\rho=6$, blue, $N_\rho=4$, green, $N_\rho=2$. }
        \label{fig:br2vr}
\end{figure*}
For relatively low stratification, the area occupied by the positive upflows and negative downflows is approximately the same at the surface. The colormap in figure~\ref{fig:br2vr}a illustrate this through the distribution of radial velocity component $u_r$, perpendicular to the surface plane, for $N_\rho=2$ and $\widetilde{Ra}=32$. As density gradient between the two spherical shell becomes more pronounced, convective cells also become more anisotropic. For $N_\rho=6$, $\widetilde{Ra}$, although the upward motions occupy larger surface, they are compensated by much stronger and localized downward sweeps (figure~\ref{fig:br2vr}b). Such asymmetry results from pressure fluctuations enhancing buoyancy driving in downflows and braking buoyancy in upflows~\citep{hurlburt1984two}. Hence, the probability of having extreme negative velocity fluctuations at the surface for $N_\rho=6$ is much larger than for $N_\rho=2$, and so the whole distribution of $u_r$ is skewed toward negative values (figure~\ref{fig:br2vr}c, in red). This does not take place at $N_\rho=2$ where the distribution of $u_r$ is much more symmetric (figure~\ref{fig:br2vr}c, in green), and probabilities of up- and down-flows are nearly equal. 

This property of the probability distribution can be characterized by the Fisher-Pearson standardized moment coefficient, 
\begin{equation}\label{eq:skewness}
    G_1 = \frac{\sqrt{k (k-1)}}{k-2} \frac{(1/k) \sum_{i=1}^k \left( u_{r,i}(r) - \overline{u_r}(r) \right)^3}{\left[ (1/k) \sum_{i=1}^k  \left(u_{r,i} (r) - \overline{u_r (r)}\right)^2 \right]^{(3/2)}} , 
\end{equation}
which serves as a measure of the asymmetry of the velocity distribution at given $r$ (here $r/r_o=0.9$). We calculated this coefficient from a series of $k$ flow snapshots for different values of $\widetilde{Ra}$ and $N_\rho$, and plotted it in figure~\ref{fig:br2vr}(d). For $N_\rho=2$, the skewness coefficient is relatively small and nearly saturates at the largest values of $\widetilde{Ra}$. With increase of stratification to $N_\rho=4$ or $6$, the left tail of the probability density distribution for $u_r$ becomes longer and strong negative downflows progressively more plausible. Here, the distribution skewness does not show signs of saturation with $\widetilde{Ra}$, indicating that surface convection will become even more anisotropic if enhanced.  Note that the mean values of $u_r$ become more positive as the skewness of its distribution increases, since the surface area of upward motions expands (figure~\ref{fig:br2vr}b).  

This process is related mathematically to development of convergent flows in stratified flows. In anelastic approximation, mass conservation equation takes the form of
\begin{equation}\label{eq:div_u}
    \nabla \cdot \left(\rho(r)\mathbf{u} \right) =0, \quad \rho(r) \nabla \cdot \mathbf{u} + u_r \frac{\partial \rho(r)}{\partial r} =0.
\end{equation}
In compressible fluids, $\nabla \cdot  \mathbf{u} \neq 0$ and is proportional to $u_r$ according to \eqref{eq:div_u}. Since density is a positive-definite function that decreases along spherical shell, $\partial \rho/\partial r<0$, convective downflows with $u_r<0$ are linked with the regions of convergent flows, $\nabla \cdot \mathbf{u} <0$. These flows are involved in the equation for magnetic energy,
\begin{equation}\label{eq:mag_energ}
 \frac{1}{2}\frac{\partial B^2}{\partial t} = -B^2 \left( \nabla \cdot \mathbf{u} \right) + \mathbf{B} \left(\mathbf{B} \cdot \nabla \right) \mathbf{u} - \mathbf{B} \left( \mathbf{u} \cdot \nabla \mathbf{B}\right) + \cdots,
\end{equation}
  where the first term reflects compression of magnetic field lines  and amplification of magnetic energy due to convergent flows. We illustrate this process with 2D joint probability density distributions for $u_r$ and the radial magnetic field $B_r$ (figure~\ref{fig:br2vr}c). The probability densities were computed by binning the data for $u_r$ and $B_r$ into $100$ and $50$ bins of equal size, respectively, in the same interval of $u_r \in [-2000,2000]$ and $B_r^2 \in [0,2000]$. 

The turbulent layer, forming at the surface for $N_\rho=6$, is associated with strong magnetic induction process near the surface, generating small-scale magnetic structures.  Similarly to $Ro_l$,  magnetic Reynolds number $Rm = u_{rms} d/\eta$ (see table~\ref{tab:sim_param}) also becomes progressively higher with increasing $N_\rho$ and $\widetilde{Ra}$, enhancing induction. As a result, the characteristic value of magnetic energy fluctuations near the surface also increases with $N_\rho$, up to $2$ times in figure~\ref{fig:br2vr}(c). For  $N_\rho=6$, strong downward motion from the negative, long tail of the velocity distribution are associated with the enhanced probability of stronger magnetic energy, unlike the upflows of the ``shorter", positive tail (figure~\ref{fig:br2vr}c, in red). This effect is much less present for $N_\rho=2$ (in green), as the energy of magnetic fluctuations is distributed together with the more symmetric distribution of velocity fluctuations. In other words, negative fluctuations of velocity field appear more correlated with the more energetic field lines  in stratified turbulence. We attribute this effect to strong anisotropic sweeps being able to compress magnetic field fluctuations induced by small-scale convection at the surface through the term proportional to $B^2 \nabla \cdot u$ in \eqref{eq:mag_energ}. This process is illustrated by the contours of  $B_r$ in figure~\ref{fig:br2vr}, which are much more aligned with the localized downflows for $N_\rho=6$ (panel b) than for $N_\rho=2$, where they are more or less evenly spread across the domain surface (panel a). This compression provides a mechanism of inward transport of magnetic energy generated by strongly turbulent, rotationally unconstrained small-scale motions in outer layer with $Ro_l>1$ (figure~\ref{fig:sim_param}a), suggesting importance of this region for the dynamo. To see how it is results in the transfer of magnetic energy to the large scales, we relate it to the global, mean-field advection of magnetic field by the flow -  magnetic pumping.

\subsection{Magnetic pumping}\label{sec:pumping}
\begin{figure*}
\centering
    \begin{subfigure}[b]{0.3\textwidth}
    \centering
        \caption{}
         \includegraphics[width=\textwidth]{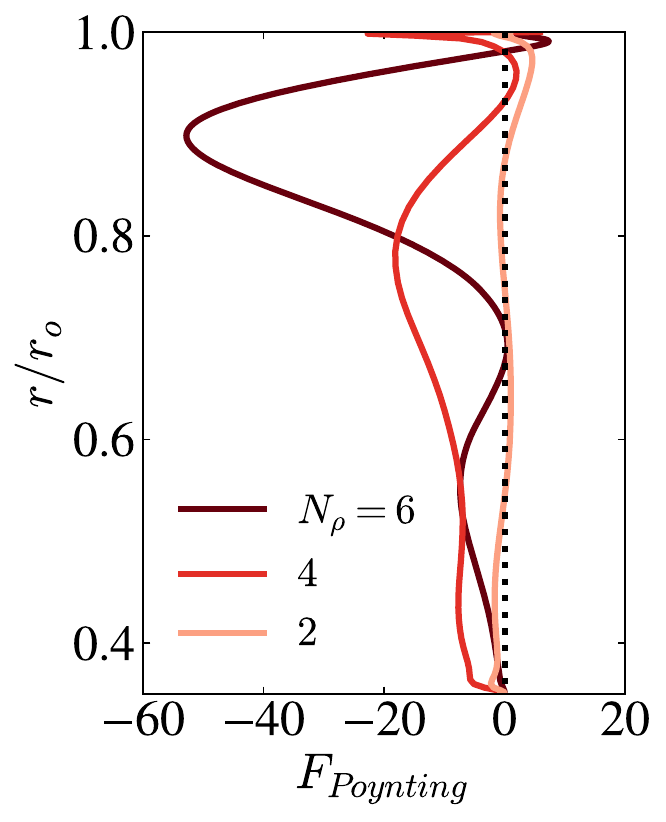}
     \end{subfigure}
     \hfill
    \begin{subfigure}[b]{0.3\textwidth}
    \centering
        \caption{}
         \includegraphics[width=\textwidth]{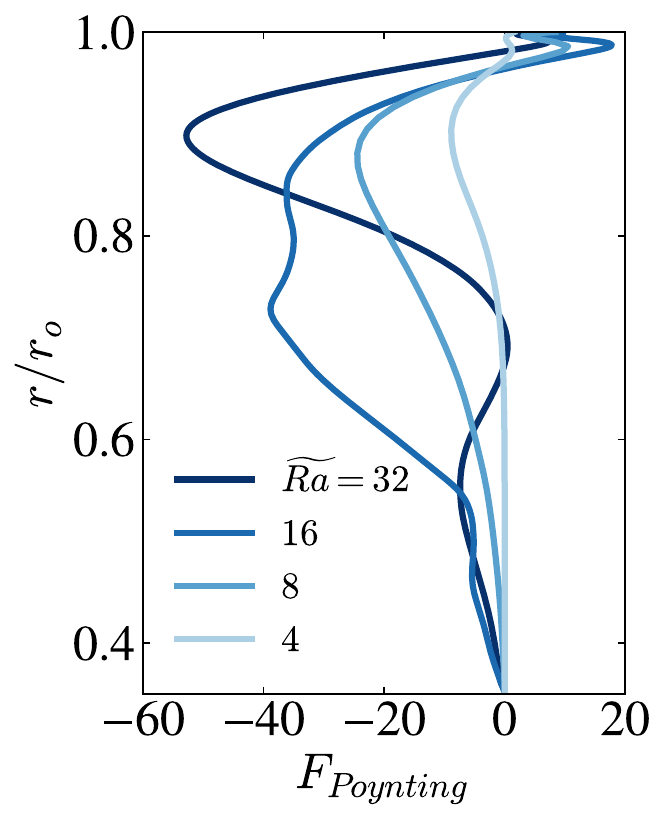}
     \end{subfigure}
     \hfill
         \begin{subfigure}[b]{0.3\textwidth}
         \centering
        \caption{}
         \includegraphics[width=\textwidth]{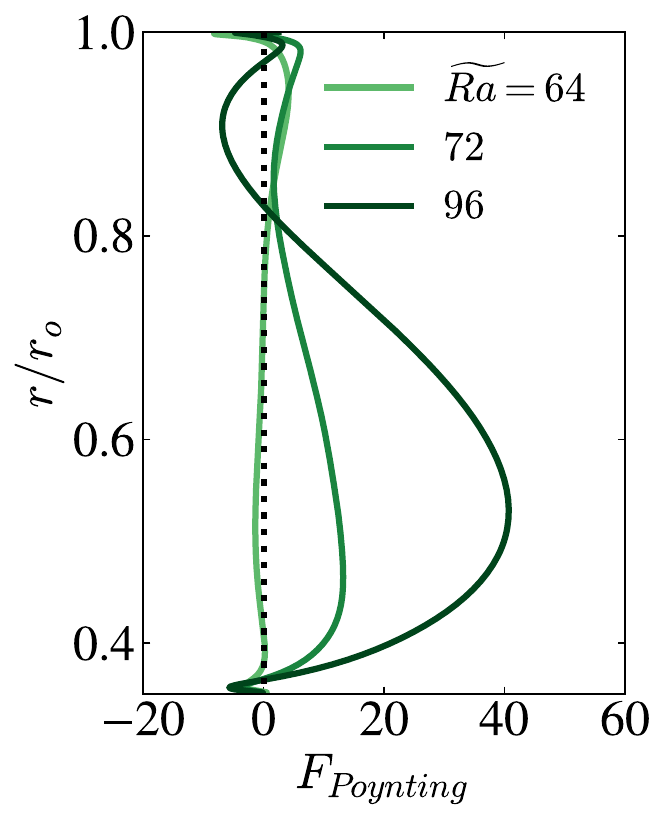}
     \end{subfigure}

             \caption{ (a) Poynting flux~\eqref{eq:poynt}, integrated in latitude $\theta$, as a function of spherical radius $r$ and $N_\rho$, $\widetilde{Ra}=32$. (b) Poynting flux as a function of $\widetilde{Ra}$, $N_\rho=6$. (c) Same but for $N_\rho=2$. }
        \label{fig:poynting_radius}
\end{figure*}
Magnetic pumping is a mechanism of large-scale magnetic field generation in the mean-field dynamo theory \citep{moffatt1978field}. Using a set of approximations, the electromotive force $\varepsilon $, generating mean-field from correlated fluctuations of velocity and magnetic field can be parametrized as 
\begin{equation}\label{eq:emf}
    \varepsilon =  \langle \bm{u} \times \bm{B} \rangle \approx \alpha\langle \mathbf{B} \rangle + \bm{\gamma} \times \langle \bm{B} \rangle + \cdots,
\end{equation}
where $\langle B \rangle$ is large-scale, axisymmetric magnetic field.
In the expression above, higher-order contributions, proportional to derivatives of $\langle B \rangle$ and corresponding to the turbulent diffusion, were omitted for simplicity.  In~\eqref{eq:emf},  $\alpha$-effect generates large-scale field through perturbing magnetic field lines by cyclonic motions, and $\gamma \times \langle B \rangle$ describes the pumping of magnetic field with a speed $\mathbf{\gamma}$. Mean-field induction equation shows that $\mathbf{\gamma}$ is an addition to the mean velocity that also acts on the mean magnetic field:
\begin{equation}\label{eq:mean_field_B}
    \mathbf{\langle \bm{B} \rangle}_t = \nabla \times \left(  \langle \bm{U}\rangle \times\langle \bm{B} \rangle  + \bm{\gamma} \times \langle \bm{B} \rangle +  \alpha \langle \mathbf{B} \rangle - \eta_T \nabla \times \langle \bm{B} \rangle \right). 
\end{equation}
In essence, magnetic pumping is a turbulence-generated component of the electromotive force~\citep{drobyshevski1977magnetic} \anna{that} can help to redistribute magnetic energy to large scales. In rotating convection, it can be viewed as a systematic transport of magnetic energy by anisotropic convective columns toward the bulk of the domain: convective structures are stretched vertically if their location is perturbed inwards, and this increases the length scales of magnetic flux tubes that are carried by them~\citep{schrinner2012dipole}. 
Equation~\eqref{eq:mean_field_B} shows that this effect is capable of generating large-scale magnetic fields. However, evaluating $\gamma$ is not straightforward. The most rigorous way is the test-field method, but it requires to solve a set of additional equations~\citep{schrinner2012dipole} for auxiliary variables. Another frequently used method is the method of snapshots, i.e. point-wise reconstruction of the tensor coefficients from time series of $\varepsilon$ and mean magnetic fields at each point in the computational domain~\citep{simard2016characterisation}. This method is more straightforward to use, yet it suffers from noise in strongly turbulent conditions like those explored here and may not always give correct results~\citep{warnecke2018turbulent}. To unambiguously quantify the transport of magnetic energy across the domain for different parameter regimes in our simulations, we use an indirect measure of this process, the Poynting flux
\begin{equation}\label{eq:poynt}
 \mathbf{F}_{Poynting} \propto \langle (\mathbf{v} \times \mathbf{B}) \times \mathbf{B}\rangle_{\phi,t} = \langle B^2 \mathbf{u}_\perp \rangle_{\phi,t},    
\end{equation}
averaged over longitude and in time. This vector describes the direction in which magnetic energy is transported by the velocity $\mathbf{u}_\perp$ perpendicular to magnetic field lines.

In figure~\ref{fig:poynting_radius} we plot the distribution of the radial component of Poynting flux with radius, integrated in latitude $\theta$, for three series of simulations in figure~\ref{fig:sim_param}b. The influence of stratification on the flux at a given $\widetilde{Ra}=32$ is relatively clear: while for $N_\rho=2$ the  flux is nearly zero throughout the domain, it becomes negative for $N_\rho=4$, reflecting the inward transport of magnetic energy toward the inner sphere (figure~\ref{fig:poynting_radius}a). For $N_\rho=6$, this flux becomes more localized, with a strong peak developing near the surface, at $r/r_o=0.9$. This flux results from consistent enhancement of turbulence near the outer shell, and becomes more and more pronounced as $\widetilde{Ra}$ grows (figure~\ref{fig:poynting_radius}b). This indicates that the energy generated and compressed by anisotropic convection in the turbulent surface layer is systematically transported into the deep interiors. On the other hand, in weakly stratified flows ($N_\rho=2$), the enhancement of convection results in magnetic flux becoming predominantly positive throughout the domain, and the outward transport of magnetic energy (figure~\ref{fig:poynting_radius}c). This is potentially related to inversion of magnetic pumping effect: energetic turbulent motions, developing near the inner sphere, expel outwards generated by them magnetic fluctuations. Although a region of negative flux develops around $r/r_o \approx 0.9$ at the outer surface for $\widetilde{Ra}=96$, it is entirely compensated by even stronger flux expulsion region at $r/r_o \in [0.35, 0.8]$. Correspondingly, the average energy of large-scale dipole fields remains low with increase of $\widetilde{Ra}$ in weakly stratified flow, while for strongly stratified cases dipolarity increases (figure~\ref{fig:dipolarity}a).

 Note that the dependence of Poynting flux on the radius is non-monotonic. For example, the flow with $N_\rho=6$, $\widetilde{Ra}=32$ case has two regions of negative flux build-up, separated by a minimum at $r/r_o \approx 0.7$. It also varies with parameter regime: unlike for $\widetilde{Ra}=32$, $\widetilde{Ra}\leq 16$, the flux is enhanced nearly throughout the whole domain. This suggests that the flux distribution is a 2D function of both $r$ and $\theta$. To clarify this point, we show the streamlines of the poloidal component of the Poynting flux, $\mathbf{F}^{P}_{Poynting} = F_r \mathbf{e_r} + F_\theta \mathbf{e_\theta}$ in figure~\ref{fig:flux_streamlines}. The corresponding magnetic regimes are described in section~\ref{sec:mag_solutions}. For $N_\rho=2$, where the integrated in $\theta$ flux is weak, the flow is separated in three different regions - inward-flux zone near the equator, outward around the mid-gap, and another inward-flux zone near the inner sphere (figure~\ref{fig:flux_streamlines}a). For $N_\rho=6$ and $\widetilde{Ra}=16$, where the large-scale coherent waves develop, Poynting flux is consistently directed inwards across the entire domain, except for a narrow zone at the equator. Unstable dipole dynamos ($N_\rho=6$, $\widetilde{Ra}=32$) develop three interchanging positive and negative regions of magnetic flux (figure~\ref{fig:flux_streamlines}c), similarly to weakly stratified case in panel (a).  But unlike $N_\rho=2$, in this case the width of the inner  region of inward-directed flux is considerably larger, and occupies more than half of the domain; there is a clear negative (positive) vertical flux of magnetic energy in upper (lower) hemisphere. This suggests more efficient transport of magnetic flux in the polar regions, similarly to flux distribution of  axisymmetric large-scale dynamo waves (figure~\ref{fig:flux_streamlines}b). Note that dipolar field component is most prominent in the polar regions inside the tangent cylinder (figure~\ref{fig:butterflys}c and also meridional slices of $\langle B_r \rangle_\phi$, not shown).

\subsection{Poynting flux and zonal flows}\label{sec:shear2flux}

Figure~\ref{fig:flux_streamlines} also demonstrates that the local direction of the Poynting flux and the effective surface from which magnetic flux is advected inside the domain is linked to the distribution of zonal flows. The distribution of magnetic flux lines for weakly stratified turbulence at $N_\rho=2$ and $\widetilde{Ra} =32$ shows that the convergent flux zone at the equator is the bulk is correlated with the area of strong shear between the prograde and the retrograde jets, where part of the Poynting flux is transformed from poloidal to toroidal (figure~\ref{fig:flux_streamlines}a). The maximum of the retrograde jet in the bulk coincides with divergent flux zone, and seemingly isolates the area of flux accumulation inside the tangent cylinder. Since the jet is located close to the tangent cylinder, the surface from which the flux accumulates is effectively small. The flows with strong stratification and weaker turbulence (e.g. $N_\rho=6$, $\widetilde{Ra}=16$), featuring large-scale oscillatory dynamos, also exhibit anti-solar rotation (figure~\ref{fig:flux_streamlines}b). However, inner retrograde flow in the bulk is very weak in this case, and so the flux is directed predominantly radially everywhere and collected from nearly all latitudes.

The flows with $N_\rho=4$, $6$, and $\widetilde{Ra} =32$, develop two different types of differential rotation: weaker three-layer and stronger anti-solar (figure~\ref{fig:flux_streamlines}c,d). Although a fraction of Poynting flux is sheared near the surface  by the interface between the equatorial retrograde jet and its neighbouring weak prograde flow during the intervals with three-layer rotation, the prograde jet is located further away from the inner sphere (figure~\ref{fig:flux_streamlines}c). The corresponding quiescent area in the bulk, where radially-directed flux accumulates, occupies the area twice as large than that of $N_\rho=2$ (figure~\ref{fig:flux_streamlines}a). In the second configuration of anti-solar rotation, Poynting flux is again directed downwards in the equatorial region of the surface retrograde jet. However, this flux is sheared at the interface between the two layers of the zonal flow and becomes less coherent within the prograde jet area. This results in  strengthening of axisymmetric toroidal component of magnetic field in the equatorial region during multi-polar, high-energy intervals (not shown here). In the meanwhile, vertical advection of magnetic energy inside the tangent cylinder is disrupted, and the large-scale dipole component is lost. For weakly stratified and strongly turbulent cases like $N_\rho=2$, $\widetilde{Ra}=96$, similar  configuration of zonal flows, accompanied by a stronger prograde jet near the tangent cylinder (figure~\ref{fig:sim_param}b, inset), results in positive sign of Poynting flux and expulsion of magnetic energy from the bulk (figure~\ref{fig:poynting_radius}c).

\section{Discussion}\label{sec:dics}
\begin{figure*}
    \centering
 \begin{subfigure}[b]{0.25\linewidth}
         \centering
        \caption{}
         \includegraphics[width=\linewidth]{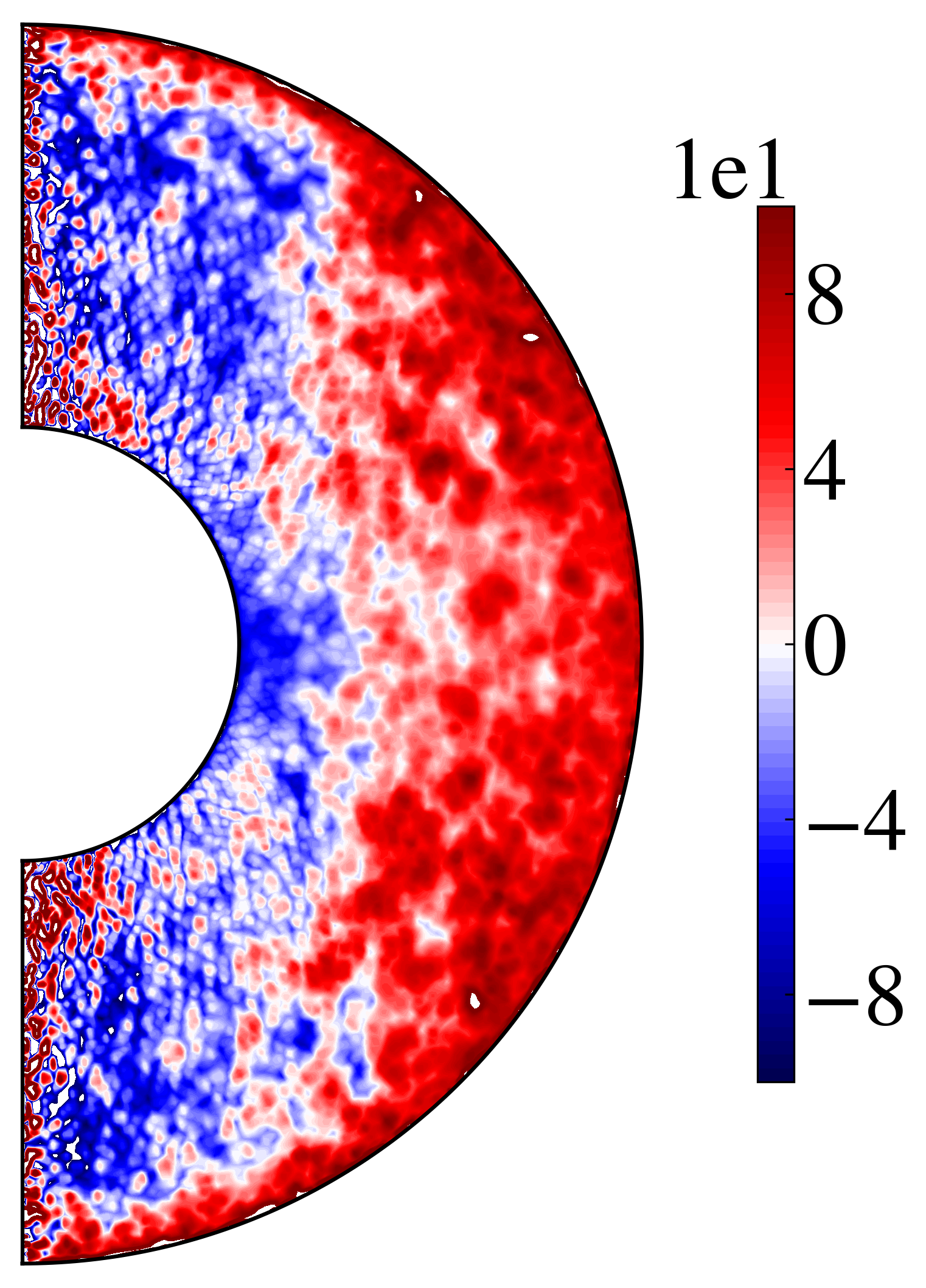}
     \end{subfigure}
\hfill 
 \begin{subfigure}[b]{0.25\linewidth}
         \centering
        \caption{}
         \includegraphics[width=\linewidth]{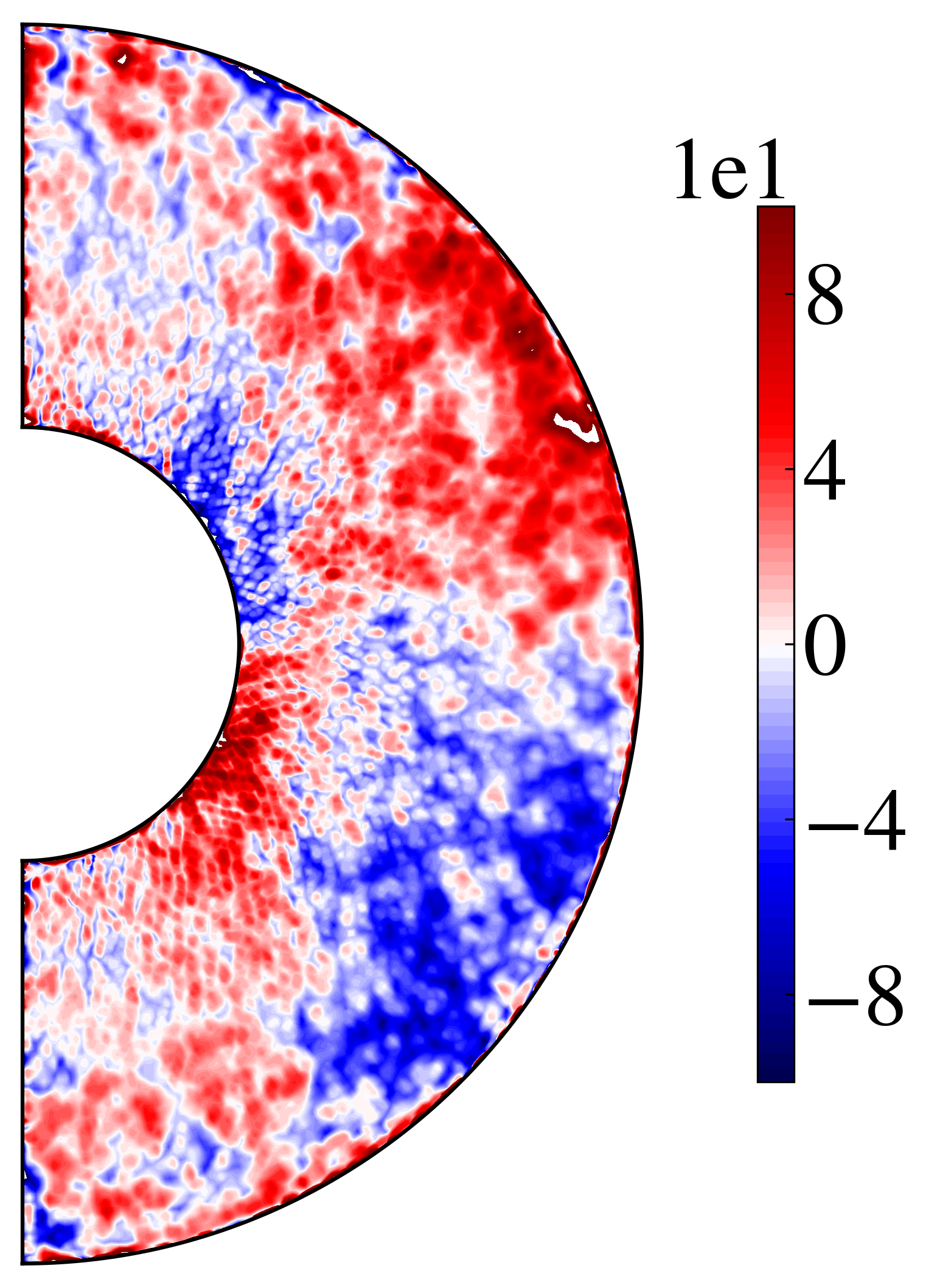}
     \end{subfigure}
\hfill
     \begin{subfigure}[b]{0.45\linewidth}
         \centering
        \caption{}
         \includegraphics[width=\linewidth]{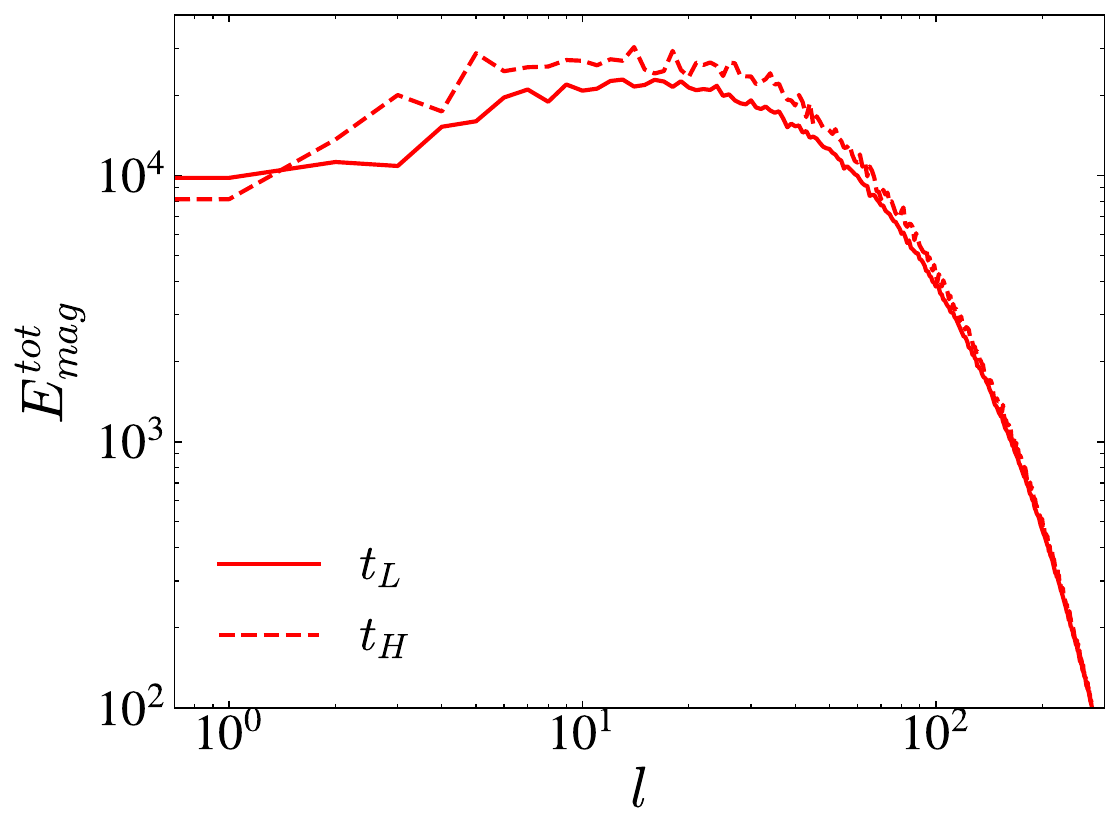}
     \end{subfigure}
    \caption{ (a,b) Components of the mean-field $\mathbf{\gamma}$-coefficient, calculated for $N_\rho=4$, $\widetilde{Ra}=32$ using method of snapshots~\citep{simard2016characterisation}. (a) Radial component, $\gamma_r$; (b) latitudinal component, $\gamma_\theta$. (c) Total magnetic energy near the surface ($r/r_o=0.9$) for run $N_\rho=6$, $\widetilde{Ra}=32$ as a function of spherical harmonic degree $l$ for high- ($t_{H}$) and low-energy ($t_L$) dynamo states.}
    \label{fig:gammas}
\end{figure*}

Recent observations show that magnetic dynamics either in form of periodic oscillations or reversals is encountered in low-mass stars~\citep{jeffers2023stellar}. While many of them have dipolar magnetic topology, other show a mix of dipolar and quadrupolar magnetic field \citep{willamo2022zeeman}. The magnetic topology and dynamics in our simulations, as given by figures~\ref{fig:butterflys} and~\ref{fig:filteredBr}, is representative of these observations. In our most strongly stratified and turbulent runs like those in figure~\ref{fig:butterflys}(c), which are closer to stellar conditions, magnetic field behaves aperiodically. The reversals and switches between dipolar and multipolar topologies through dipole decay and build-up take place on a time scale of $50-100$ rotation units. With the average solar rotation rate of 28 days as a reference, reversals of magnetic field in our model would take place on the timescale of 4-8 years, compatible with spectropolarimetric observations. Note that to make a one-to-one comparison with ZDI magnetic maps, one needs to take into account inclination of rotation axis and the number of observations per rotation period of a star \citep{hackman2024convective}. 

Our analysis of correlation between radial velocity and magnetic field at the surface, together with enhancement of inward-directed Poynting flux, indicates that magnetic pumping mechanism is responsible for development of such large-scale axisymmetric magnetic states in strongly stratified turbulence. Development of turbulent, rotationally unconstrained thin surface layer at $N_\rho=4$ and $N_\rho=6$ (figure~\ref{fig:sim_param}b), accounting for $10-20$\% of the entire domain, is a natural outcome of such convection. Such a layer can contribute to both induction of magnetic fluctuations, and to transport of this energy inwards (figure~\ref{fig:br2vr}c). We find that the increase in dipolarity with $N_\rho$ and $\widetilde{Ra}$ in our simulations (figure~\ref{fig:dipolarity}) is correlated with enhancement of inwardly-directed magnetic Poynting flux (figure~\ref{fig:poynting_radius}a,b), as opposed to low-stratified turbulent cases where magnetic energy is expelled outwards (figure~\ref{fig:poynting_radius}c). These results are in agreement with previous studies of magnetic pumping effect in Cartesian geometry, for example with~\citet{tobias1998pumping}, who found that  strong downflows in stratified convection can advect magnetic field sheets initially located at the top of convection zone  \anna{downwards}, thereby concentrating magnetic energy at radiative-convective interface. This effect is also considered one of the main ingredients in solar dynamo theory, as it is able to counteract magnetic buoyancy bringing magnetic flux tubes to the surface~\citep{brandenburg2005case}. More recent studies in global spherical geometry also found compression to be an important mechanism for generation of strong magnetic fields, which enforce solar-like prograde rotation at the equator through the magnetic transport of momentum~\citep{hotta2022generation}. 

Magnetic pumping coefficient $\mathbf{\gamma}$ is a velocity-like vector and therefore magnetic pumping in fact three-dimensional effect~\citep{ossendrijver2002magnetoconvection}.  To confirm its presence in our simulations,  in figure~\ref{fig:gammas}(a,b) we show the radial and the latitudinal components of $\mathbf{\gamma}$, as derived in equation~\eqref{eq:emf}, calculated using the method of snapshots~\citep{simard2016characterisation} on the data from dipolar solution at $N_\rho =4$, $\widetilde{Ra}=32$. The radial component of $\gamma$-effect is negative in the areas inside the tangent cylinder at high latitudes, indicating overall inward advection of magnetic field in these areas (figure~\ref{fig:gammas}a). On the other hand, in the outer, equatorial region of strong shear, $\gamma_r$ is positive, indicating outward pumping. The latitudinal component, $\gamma_\theta$, is anti-symmetric with respect to the equator, with reflects equivalent converging flows in the equatorial plane and divergent flows in deep interiors (figure~\ref{fig:gammas}b). These results are consistent with manifestation of the large-scale field predominantly at high latitudes (figure~\ref{fig:butterflys}). Moreover, the spatial distribution of $\gamma_r$ and $\gamma_\theta$ is compatible with previous results for stratified convection leading to dipolar solutions \citep{schrinner2012dipole}, which suggests that axisymmetric magnetic pumping mechanism, induced by the combination of stratification and strong turbulence at the surface, is indeed responsible for the encountered in simulations dipoles. Such mechanism could be also responsible for the dipolar stellar-like dynamo with co-existing small- and large-scale fields at the surface reported in~\cite{yadav2015explaining}, although with much smaller values of $Ro_l \leq 0.04$ and thus more rotationally constrained. As $N_\rho$ is increased, $\gamma$ obtained with the method of snapshots becomes considerably more noisy yet its structure is similar; we leave its detailed analysis for the future work.

Another measure of spatially distributed transport of electromagntic energy, meridional distributions of Poynting flux, is vertically aligned due to the system rotation and depends on the distribution of zonal flows (figure~\ref{fig:flux_streamlines}), suggesting that differential rotation plays an important role in  re-distributing the flux across the domain. Stronger zonal flows in the area of tangent cylinder prevent magnetic flux from accumulating there and result in non-axisymmetric solutions with smaller scales (figure~\ref{fig:flux_streamlines}a,d). Weakened zonal flows in this area are accompanied by more axisymmetric magnetic fields, either quadrupolar or dipolar (figure~\ref{fig:flux_streamlines}b,c).  In agreement with figure~\ref{fig:gammas}, more quiescent polar zones seems to be the locations where magnetic energy transport is most prominent (figure~\ref{fig:flux_streamlines}b,c).  In all these cases, the distribution of Poynting flux is strongly constrained by zonal flows, with the regions of strong differential rotation acting as radial flux attractors and transforming it into poloidal flux and fields. 

The transition between periodic large-scale waves and dipolar solutions has been attributed previously to increase of the period of oscillatory dynamo solutions due to fluctuations in $\alpha$-effect in simplified one-dimensional mean-field dynamo models~\citep{Richardson2010EffectsO}. We suppose that similar mechanism can be responsible for destabilization of  the periodic large-scale waves at  $N_\rho=4$ and $6$ with increase of $\widetilde{Ra}$, although a rigorous analysis of temporal fluctuation of $\alpha$-effect and the corresponding mean-field equation is needed to confirm this in the future. In our simulations, such non-oscillatory dipolar solutions are unstable and exhibit either reversals of polarity or stochastic transitions to a multi-polar state with higher energy, as  described in section~\ref{sec:cycle}. We interpret this unstable behavior as a predator-prey interaction of zonal flow and magnetic field: the system attempts to escape to a state with strong anti-solar rotation and strong turbulent magnetic fields, while increasing magnetic tension from these fields pushes it back to the original state with weaker, dipolar magnetic field and weaker three-layer rotation. This is particularly visible from the phase space map in figure~\ref{fig:phase_space}(c), with flow trajectories colored by the Lorentz force. Such predator-prey transitions have been observed before both in hydrodynamic convection \citep{busse2006parameter} and in dynamo simulations as Type 2 modulation producing minima in magnetic activity~\citep{raynaud2016convective}. We suggest that such quasi-stable flow states result from the flow being at the verge of transition between anti-solar and solar rotation.   While magnetic fields tend to postpone such transition compared to pure hydrodynamic flows \citep{fan2014simulation,karak2015magnetically}, such transition nevertheless takes place in the range of about $\widetilde{Ra} \in [70-80]$  in our magnetohydrodynamic simulations for the least numerically demanding regime of $N_\rho=2$ (figure~\ref{fig:sim_param}b). Development of the strongly turbulent shear layer with $Ro_l \leq 1$ at the surface   is correlated with the transition between solar- and anti-solar rotation for all explored $N_\rho$, which suggests that the presence of this layer can induce this transition. We expect thus that with further enhancement of convective forcing the three-layer differential rotation  (figure~\ref{fig:flux_streamlines}c) would be lost and anti-solar  state would become the only solution.  Yet the aperiodic switches between the two different shear topologies in our aperiodic dipole solutions suggest our parameter regime can be close to bi-stability of magnetically controlled transition between solar- and anti-solar rotation, essentially for a stronger density contrast than $\rho_i/\rho_o =12$ reported by \cite{karak2015magnetically}, who did not find bi-stable behavior. The three-layer structure of differential rotation, obtained thanks to the high $N_\rho$, is an interesting new result that resembles the near-surface shear layer in the Sun~\citep{vasil2024solar}; in the future, we will extend our study to even higher $N_\rho$ to study the stability of this structure with $\widetilde{Ra}$ and magnetic field. The influence of the latter can be particularly important, since magnetic field impacts both alignment or non-alignment of the flow with the axis of rotation \citep{pinccon2024coriolis}, and  the sign of differential rotation \citep{brun2022powering}.

Finally, we show the energy spectra during low- and high-energy states in the outer region ($r/r_o = 0.9$) for $N_\rho=6$, $\widetilde{Ra}=32$, as a function of spherical harmonic degree $l$ (figure~\ref{fig:gammas}c). The spectra are in agreement with the time series of magnetic energy for the two flow regimes in figure~\ref{fig:phase_space}, with energy decreased across all spherical harmonics during quasi-dipolar state. The $l$-spectra in this regime shows a certain degree of separation between the large and the small scales, with a peak on small scales around $l=20$ and build-up of energy at the dipolar mode with $l=1$. It is this build-up that is observed as enhanced dipolarity, when magnetic field is filtered at the surface. During the high-energy state, all spherical harmonics but the smallest ones receive additional amount of energy, so that the spectrum fills up and does not show anymore the signs of scale separation. It is possible that this scale separation appears due to spatial anisotropy of stratified convection, separating generation of small-scale magnetic fluctuations at the surface from large-scale fields generated through magnetic pumping. However, even stronger density stratification is necessary to see whether the trend of scale separation becomes more pronounced; we leave exploring this idea in depth for the future work.

\section{Conclusions}\label{sec:conc}

Previous studies of stratified convective dynamos by~\cite{gastine2012dipolar} and~\cite{raynaud2015dipolar} reported dipole collapse with increase in density stratification, exploring  parameter regimes with lower values of $\widetilde{Ra}$ and $N_\rho\leq 3$ than those explored in this work. Essentially, these dipolar dynamos correspond to classic steady low-$Ro$ dynamos, and are quickly destabilized by inertia when $Ro \sim 0.1$~\citep{christensen2006scaling}. In this work, we explored more stratified and turbulent models, up to $N_\rho=4,6$ and $\widetilde{Ra}=32$,  which resulted in larger integral values of $Ro>0.1$ (table~\ref{tab:sim_param}).

In all our models, magnetic fields feature broad spread of energy across spherical harmonics, with a lot of small-scale magnetic fluctuations at the surface  (figure~\ref{fig:gammas}c). According to the classic definition of dipolarity as the ratio between the energy of the dipole and the total magnetic energy, used by \cite{gastine2012dipolar}, such fields would be classified as multipolar and small-scale. Here, we compared the strength of the dipolar mode with  the energy of up to $7-11$ large-scale spherical harmonics (equation~\ref{eq:fdip}). 
 The strength of dipolar component is consistently increasing with both stratification and turbulence vigor, which suggests similar magnetic regimes are possible in low-mass stars, frequently featuring dipolar magnetic fields. On the large scales, our models reproduce magnetic topologies commonly observed in these stars - multipolar, oscillatory and dipolar. They also reproduce main dynamical ingredients of such fields, such as dipolar reversals, and transitions between dipolar and multipolar regimes. With the averaged \anna{solar} rotation rate as a reference, reversals of magnetic field in our model would take place on the timescale of 4-8 years, compatible with spectropolarimetric observations of stellar cycles \citep{jeffers2023stellar}.  We argue that this large-scale definition of dipolarity~\eqref{eq:fdip}, defined up to a suitable low-order truncation threshold $l_{cut}$, better characterizes magnetic topology in stellar dynamo models, given that large-scale fields detected by spectropolarimetry are \anna{an order of magnitude} smaller than total fields from stellar photometric observations.

We identify that one of the direct consequences of increasing both stratification and turbulence is the development of the turbulent surface layer with surface $Ro_l>1$ and weakened influence of the Coriolis force, while the deep interiors of the flow become more and more rotationally constrained (figure~\ref{fig:sim_param}a). Strong downflows promoted by the anisotropic convection of the turbulent surface layer, together with stratified and more quiescent convective columns in the flow interiors, induce an axisymmetric magnetic pumping mechanism which scales with $N_\rho$ and $\widetilde{Ra}$.  This physical mechanism could be relevant for dominant magnetic mode selection in other irregular magnetic cycles previously reported in slowly rotating stellar models of \cite{yadav2016magnetic,strugarek2018sensitivity,viviani2018transition} at lower stratification.  Our results suggest that magnetic pumping consistently enhances dipolar component in our simulations, to the point where prolonged dipolar periods of order $500-1000$ rotation times are observed for $N_\rho=4$ and $6$, $\widetilde{Ra}=32$. These runs are on the verge between solar- and anti-solar transition, and exhibit dynamical interaction of magnetic field and the zonal flows, transitioning between low-energy, dipolar state with weakened three-layer rotation and high-energy, multipolar states with stronger anti-solar rotation.  As in~\cite{brun2022powering}, we observe that the Lorentz force can change configuration of the differential rotation  in simulations where $Ro_l>1$; this phenomenon is favored when $N_\rho$ is high. These dynamical interactions suggest additional observational constraints based on relation between stellar fields and internal differential rotation. Such constraints could be obtained from joint  spectropolarimetry and asteroseismology studies. 

Is yet to be shown how magnetic pumping effects extrapolate to higher values of  $Ra$ and $N_\rho$ and whether dipolar solutions are constrained by transition between solar-antisolar rotation. However, the enhancement of large scales, together with emergence of surface dipole and quadrupole fields in strongly turbulent convection suggests a new framework for modeling dynamical magnetic effects in the DNS models of stellar convection. This framework supposes seeking models and parameter regimes that (i) capture dynamics of large magnetic field scales and (ii) capture the energy balance between large- and small scales that depends on the depth.  Such models will be the aim of our future work. Although magnetic dipoles are not steady states in our simulations, their existence at the interface of oscillatory quadrupolar and non-coherent multipolar solutions indicates a parameter path for further exploration of such states in stellar dynamos, extremely expensive numerically in these parameter regimes.

 \begin{acknowledgements} 
 \anna{The authors acknowledge financial support from the French program ’PROMETHEE’ (Protostellar Magnetism: Heritage vs Evolution) managed by Agence Nationale de la Recherche (ANR), and would like to thank E.~Alecian and the PROMETHEE team for inspiring discussions that helped to shape this work.  The authors are also grateful to F.~Daniel and S.~Tobias for their insightful comments, and the anonymous reviewer for their constructive feedback. This work was supported by the "Action Thématique de Physique Stellaire" (ATPS) of CNRS/INSU PN Astro co-funded by CEA and CNES, and the HPC facility MesoPSL for the computational resources.}
 \end{acknowledgements}

\bibliographystyle{aa} 
\bibliography{dynamo.bib}
\appendix
\section{Anelastic model}\label{sec:app_eqn}
In this work, we use LBR anelastic approximation, as described by~\cite{braginsky1995equations,lantz1999anelastic,jones2011anelastic}. The reference thermodynamic state of the system is assumed to be close to adiabatic hydrostatic equilibrium, and follows the polytropic structure for pressure, density and temperature, 
\algn{
P=P_{\rm c} \varw^{n+1}\; ,~~~~\rho = \rho_{\rm c} \varw^n\; ,~~~~ T = T_{\rm c} \varw \; ,
}
with
\algn{
&\varw=c_0+\frac{c_1d }{r}\; ,~~~~c_0=\frac{2\varw_{\rm o}-\chi-1}{1-\chi}\; ,\\
&c_1=\frac{(1+\chi)(1-\varw_{\rm o})}{(1-\chi)^2} \; , ~~~~\varw_{\rm o} =\frac{\chi+1}{\chi \exp(N_\rho/n)+1} \; ,
}
with reference values $T_c$, $\rho_c$ and $P_c$ at the mid-gap between the sphere boundaries, and $\xi = r_i/r_o$. 

In this approximation, we solve three equations for velocity field $\mathbf{u}$, magnetic field $\mathbf{B}$ and the entropy of the flow $S$:

\algn{
\derivp{\vec{u}}{t} + \vec{u}\cdot \vec{\nabla} \vec{u} =&  - \frac{Pm}{E} \vec{\nabla}\left( \frac{P^\prime}{\varw^n}\right) + \frac{Ra Pm^2}{Pr} \frac{S}{r^2} \vec{e}_r -\frac{2Pm}{E} \vec{e}_z \times \vec{u} \nonumber \\
&+ \frac{Pm}{E\varw^n} (\vec{\nabla}\times \vec{B})\times \vec{B}+Pm\vec{F}_\nu \; , \label{moment}\\
\derivp{S}{t} + \vec{u} \cdot \vec{\nabla} S =&\frac{Pm}{\varw^{n+1} Pr} \vec{\nabla} \cdot \left(\varw^{n+1} \vec{\nabla} S \right) \nonumber \\
&+ \frac{Di}{\varw} \left[\frac{(\vec{\nabla}\times \vec{B})^2}{E \varw^{n}} +Q_\nu \right]\label{chaleur}\\
\derivp{\vec{B}}{t} =& \vec{\nabla}\times\left(\vec{u}\times\vec{B} \right)+ \vec{\nabla}^2 \vec{B} \label{induction}\\
\vec{\nabla}\cdot \vec{B} =&0 \label{flux B}\\
\vec{\nabla} \cdot \left(\varw^n \vec{u} \right) =&0 \label{flux m}\; .
}

Here $P^\prime$ denotes the pressure perturbation of the reference state. The viscous force $\vec{F}_\nu$ can be expressed $\vec{F}_\nu= \varw^{-n} \vec{\nabla} \mathsf{S}$, where $\mathsf{S}$ is the rate of strain tensor,
\algn{
\mathsf{S}_{ij}=2\varw^{n} \left( e_{ij}-\frac{1}{3} \delta_{ij} \vec{\nabla} \cdot \mathbf{u} \right)\; , ~~~~e_{ij}=\frac{1}{2} \left(\derivp{u_i}{x_j}+\derivp{u_j}{x_i} \right) \; .
}
The dissipation parameter $Di$ and the viscous heating  $Q_\nu$are given by
\algn{
Di=\frac{c_1 Pr}{Pm Ra}\; , ~~~~Q_{\nu} = 2 \left[e_{ij}e_{ij}-\frac{1}{3} (\vec{\nabla}\cdot\vec{u})^2 \right] \; .
}

 Equations above were scaled with the shell width $d=r_o-r_i$, magnetic diffusion timescale $d^2 /\eta$, by entropy difference $\Delta S = S(r_i) - S(r_0)$, pressure scale $\Omega \rho_{\rm c} \eta$, density scale $\rho_{\rm c}$, temperature by $T_{\rm c}$, and magnetic field scale $\sqrt{\Omega\rho_{\rm c} \mu \eta}$, where $\mu$ is the magnetic permeability. Resulting dimensionless parameters $Pm$, $Pr$, $E$, and $Ra$ are described in the text in section~\ref{sec:methods}.

\section{Simulation parameters}\label{sec:app_sim_param}

\begin{table*}[h!]
    \centering
    \begin{tabular}{|c|c|c|c|c|c|c|c|c|}
    \hline
        Label & $\widetilde{Ra}$ &  $Ra^*_F$ & $Ro$ & $Re$ & $Rm$  & $Nu$ & $E^{l\leq 7, m=0}_{mag}/E^{l\leq 7}_{mag}$ & $E^{m=0}_{mag}/E^{tot}_{mag}$ \\ \hline
         \multicolumn{9}{|c|}{$N_\rho=2$, $Ra_{cr}=1.483 \cdot 10^5$}\\ \hline
         \verb|n21|  & 32 &  $2.32 \cdot 10^{-3}$ & 0.615 & 517 & 1035 & 18.1 & 0.146& 0.043 \\ \hline
          \verb|n22| & 64 & $7.31 \cdot 10^{-3}$ & 0.955  &816 &1633 &28.5 & 0.102 & 0.0287\\ \hline
          \verb|n23| & 72 &   $7.75 \cdot 10^{-3}$ & 0.971 & 865 & 1731&  26.9 & 0.147 & 0.0294\\ \hline
          \verb|n24| & 96 &$12.82 \cdot 10^{-3}$ & 1.01 & 1224& 2448& 33.4 & 0.137 & 0.044\\ \hline
    \multicolumn{9}{|c|}{$N_\rho=4$, $Ra_{cr}=4.224\cdot 10^5$}\\ \hline
        \verb|n41|& 16 &  $2.74 \cdot 10^{-3}$ & 0.393 & 528& 264& 9.2 & 0.254 & 0.091\\ \hline
        \verb|n42| & 32 &$9.28 \cdot 10^{-3}$  & 0.664 & 459 & 918 &  15.6 & 0.150 & 0.028\\ \hline
     \multicolumn{9}{|c|}{$N_\rho=6$, $Ra_{cr}=6.898 \cdot 10^5$}\\   \hline
       \verb|n61|& 4 &   $0.13 \cdot 10^{-3}$ & 0.057 & 28.4 & 56.8 & 1.76 & 0.03 & 0.037\\ \hline
       \verb|n62|& 8 &  $0.41 \cdot 10^{-3}$ & 0.116 &57.6 & 115.3&  2.75 & 0.597 & 0.264\\ \hline
       \verb|n63|& 16 &  $1.05 \cdot 10^{-3}$ & 0.181 & 98&197 &  3.5  & 0.437 & 0.202\\ \hline
       \verb|n64| & 32 & $3.76 \cdot 10^{-3}$  & 0.295 & 352 & 704  &  6.3 & 0.293 & 0.071\\ \hline
                 
    \end{tabular}
    \caption{A summary of simulations in this paper and their parameters, averaged over time where necessary. The run  \textit{n61} with $\widetilde{Ra}=4$ near convection onset corresponds to weakly developed turbulence and has a strongly non-axisymmetric large-scale magnetic field, concentrated in the southern hemisphere and was not discussed here in detail.}
    \label{tab:sim_param}
\end{table*}

In stratified convection, both the critical Rayleigh number $Ra_{cr}$, and the azimuthal wavenumber of the dominant convective mode, increase with density contrast~\cite[see e.g.][]{raynaud2018gravity}. The higher is density stratification, the higher is the numerical resolutions needed to resolve all the flow scales. In table~\ref{tab:sim_param} we give the values of $Ra_{cr}$ employed in our simulations, that allow together with section~\ref{sec:methods} and figure~\ref{fig:sim_param}b reconstruct the whole parameter space spanned by simulations in this work, together with several diagnostic quantities of each run. $Ra_{cr}$ denotes the critical value of $Ra$ for the onset of convection, $\widetilde{Ra} = Ra/Ra_{cr}$ denotes the distance from the onset, and $Ra^*_F = Ra Nu Ek^3/Pr^2$ denotes modified flux-based Rayleigh number.  $Nu$  is defined as the ratio between the luminosity of the convective model, where integral is taken over the top spherical boundary, and the basic state luminosity of conductive state,

\begin{equation}
    Nu = \frac{\int_S - \kappa \rho T \frac{\partial S}{\partial r} r^2 \sin{\theta} d\theta d\phi}{4 \pi n c_1 \xi_i^n \left( \exp N_\rho -1 \right)^{-1} } 
\end{equation}

In addition, we give the Rossby number $Ro = u^I_{rms} /(l^I_{conv} \Omega)$, based on the integrated over $r$ convective length scale and rms velocity in equation~\eqref{eq:rossby}, as introduced by~\cite{christensen2006scaling}. The Reynolds number and magnetic Reynolds numbers are defined as $Re = u^I_{rms} d/\nu$ and $Rm = Pm Re$, respectively, and  represent kinetic energy characteristics of each run. Finally, we present the energetics for magnetic field of the runs in the last two columns of table~\ref{tab:sim_param}, measured as the ratio of large-scale axisymmetric components $E_{mag}^{l\leq7, m=0}$ and the energy contained in all large scale harmonics $E_{mag}^{l\leq7}$, similarly to section~\ref{sec:cycle2dip}. In the last column of table~\ref{fig:sim_param} we compare the ratio between the axisymmetric and the total magnetic energy. These are typically order of magnitude smaller compared to the large-scale estimates, except for the runs \textit{n41}, \textit{n62}, \textit{n63} which feature energetic quadrupolar dynamo waves described in section~\ref{sec:waves_sol}. 

The difference between weakly stratified (section~\ref{sec:multi_sol}) and strongly stratified oscillatory solution (section~\ref{sec:waves_sol}) is that the latter contains much more energy in the axisymmetric modes, and shows approximately the same proportion of axisymmetry with respect to both large-scale and total magnetic energy (runs \textit{41}, \textit{n62}, \textit{n63}). On the other hand, the weakly stratified oscillatory solutions show (runs \textit{n21-24}) have much weaker axisymmetric component, especially in comparison to the total magnetic energy. Note that axisymmetric energy decreases in proportion to the total magnetic energy for more stratified runs ($N_\rho=4$ and $6$). This is due to the development of surface layer generating small-scale, non-axisymmetric magnetic fluctuations, which are then factored in total energy estimates. The time-averaged characteristics in table~\ref{tab:sim_param} also do not take into account change in energy as the dynamo undergoes dipolar-multipolar transitions.


\end{document}